\documentclass[amsmath,amssymb,reprint,bibnotes,superscriptaddress]{revtex4-2}
\usepackage{graphicx}
\usepackage{bm}
\usepackage{amsmath} 
\usepackage{graphics}
\usepackage[dvipsnames]{xcolor}
\usepackage[mathlines]{lineno}
\usepackage{braket}
\usepackage[compact]{titlesec}
\usepackage{float}

\begin{document}

\title{QUPERS: Quantum Enhanced Raman Spectroscopy}

\author{Özge Demirtas}
\affiliation{{Micro and Nanotechnology Program, Middle East Technical University, 06800 Ankara, Turkey}}
\author{Taner Tarik Aytas}
\affiliation{{Department of Physics, Akdeniz University, 07058 Antalya, Turkey}}
\author{Duygu Gümüs}
\affiliation{{Department of Chemistry, Middle East Technical University, 06800 Ankara, Turkey}}
\author{Deniz Eren Mol}
\affiliation{{Institute of Nuclear Sciences, Hacettepe University, 06100 Ankara, Turkey}}
\affiliation{{Division of Nanotechnology and Nanomedicine, Hacettepe University, 06100 Ankara, Turkey}}
\author{Batuhan Balkan}
\affiliation{{Department of Physics, Middle East Technical University, 06100 Ankara, Turkey}}
\author{Emren Nalbant}
\affiliation{{Department of Chemistry, Middle East Technical University, 06800 Ankara, Turkey}}
\author{Mehmet Emre Tasgin}
\email{metasgin@hacettepe.edu.tr}
\affiliation{{Institute of Nuclear Sciences, Hacettepe University, 06100 Ankara, Turkey}}
\affiliation{{Division of Nanotechnology and Nanomedicine, Hacettepe University, 06100 Ankara, Turkey}}
\author{Ramazan Sahin}
\affiliation{{Department of Physics, Akdeniz University, 07058 Antalya, Turkey}}
\affiliation{{Türkiye National Observatories, TUG, 07058, Antalya, Turkey}}
\author{Alpan Bek}
\email{bek@metu.edu.tr}
\affiliation{{Micro and Nanotechnology Program, Middle East Technical University, 06800 Ankara, Turkey}}
\affiliation{{Department of Physics, Middle East Technical University, 06100 Ankara, Turkey}}

\date{\today}

\begin{abstract}

{We present experimental demonstration of a recently predicted path interference phenomenon [Nanophotonics 7, 1687 (2018)]. A SERS process becomes resonant when both incident and the converted frequencies match with two plasmon resonances ---a condition which is hard to satisfy experimentally. Yet, presence of a quantum dot~(QD) coupled to the lower-energy plasmon mode has been predicted to introduce interference effect in the conversion paths and to bring the SERS process into resonance. Here, we experimentally demonstrate a 7-fold additional enhancement factor~(EF) {\it multiplying} the conventional SERS enhancement owing to the presence of QDs in addition to metal nanoparticles~(MNPs). We eliminate influence of alternative enhancement mechanisms and show that observed additional EF of 7 is solely because of the path interference~(Fano) effect taking place in the presence of the QDs. As the most significant evidence: we observe that when QD/MNP ratio exceeds $1$, EF starts to decrease. Two or more QDs sitting at different~(random) positions on a nanostar MNP surface starts to degrade the path interference effect. Even though the theoretically predicted maximum EFs of $\sim$100--1000 when QD is in the MNP hotspot are not always experimentally feasible in stochastic nature of large area SERS substrate, still, observation of a 7-fold overall increase in such a setting is significant for fabricating Raman-based sensor applications. More importantly, our scheme allows for designing selective SERS substrates. As the QD resonances can actively be shifted by external factors, such as by applying an electric field (Stark effect), it provides a previously-not-existing tool: active selective enhancement of Raman bands via an applied voltage.
}
\end{abstract}

\maketitle

\section{Introduction}

Metal nanostructures~(MNSs) localize incident field into nm-sized hot spots which makes phenomena such as single-molecule-detection~\cite{snom_single_molecule} and near-field optical microscopy \cite{ovali_single_APL} within reach. Hotspots enhance nonlinear processes even more strongly. This is because nonlinear processes take place over plasmon modes~\cite{grosse2012nonlinear} where both incident and converted fields are localized at the same hotspot. This results in extremely large overlap integrals~\cite{bookchapter,noor2020mode} that can enhance, e.g., Raman process by 10 orders of magnitude~\cite{nie_single_molecule} ---surface enhanced Raman scattering (SERS). SERS is an indispensable spectroscopy technique as each molecule has its own vibrational finger print.  

A nonlinear process ---in particular SERS--- possesses its maximum intensity (resonance) when both incident ($\omega$) and converted ($\omega_{\scriptscriptstyle R}$) frequencies coincide with two plasmon modes ($\Omega$, $\Omega_{\rm \scriptscriptstyle R}$), respectively. However, limitations in nanofabrication techniques do not always render the alignment of both fields feasible. This may cause, for example, the converted field $\omega_{\scriptscriptstyle R}$ not to be resonant with the second plasmon mode $\Omega_{\rm \scriptscriptstyle R}$.

Recent theoretical studies ---employing both analytical and numerical solutions of 3D Maxwell equations--- have shown that in this case, SERS process can still be brought into resonance by help of an auxiliary~(aux) quantum object~(QO), such as a quantum dot~(QD), coupled to the $\Omega_{\rm \scriptscriptstyle R}$ plasmon mode~\cite{postaci_silent_2018,gencaslan2022silent}. This phenomenon ---which appears also in other nonlinearities--- is the plasmon-analog of nonlinearity control observed 3-4 decades ago in quantum optics experiments with atomic vapors~\cite{schmidt1996giant}. They are also referred to as Fano resonances (FRs) in the nonlinear response~\cite{paspalakis2014strongly,singh2016enhancement}.

It is demonstrated on a simple analytical expression that presence of the QO introduces an extra term which can cancel the off-resonant $(\Omega_R-\omega_R)$ term~\cite{postaci_silent_2018}.
 Thus, presence of the aux QO can result in an increase in the SERS enhancement factor~(EF). This phenomenon is called as path interference effect referring to the former studies in quantum optics~\cite{schmidt1996giant,ScullyZubairyBook}.  In our earlier work~\cite{postaci_silent_2018} we have demonstrated this interference effect both analytically and by solving Maxwell equations.
 
This phenomenon has an invaluable potential, especially for use in post-fabrication tuning of SERS substrates, since only a few alternative mechanisms providing control over their spectral response  exists at the present~\cite{PS_plasmon_tuning}. By using the method utilized in this work, it is possible to {\it selectively enhance} target Raman bands of specific analytes over Raman bands of the non-specific molecules forming the specimen matrix~\cite{PS_nanostars}. This, in turn, is expected to have a high impact on efficient sensing of trace molecules, such as environmental stress factors \cite{paresh_2011} or  chemical warfare agents \cite{numan_2022}. This is because selective enhancement of indicative specific Raman bands can significantly improve  signal-to-noise ratio~(SNR) in detection. Moreover, electrical control~\cite{flatte2008giant,empedocles1997quantum,asif2024voltage,gunay2023demand} of the QO level-spacing~($\Omega_{\rm \scriptscriptstyle QO}$) may allow fine tuning in the integrated SERS devices. By electrical-tuning of $\Omega_{\rm \scriptscriptstyle QO}$, one can actively vary the enhancement of different Raman bands in the same sample.  

In this paper, we report the first experimental observation of quantum enhanced SERS~(QUPERS). In line with our theoretical findings, an aux QD ---whose level-spacing $\Omega_{\rm \scriptscriptstyle QD}$=562 nm is quite off-resonant with the pump ($\omega$=660 nm) and Raman-converted ($\omega_{\scriptscriptstyle R}$=715 nm) frequencies--- is found to enhance the SERS signal by 7 times~\cite{PS_3times_enhancement}. This enhancement, which \textit{multiplies the field enhancement}, is significantly strong as QDs are distributed stochastically (randomly) in which most of them do not necessarily reside at the hotspots. Our MNPs are star-shaped gold nanoparticles (gold nanostars ---AuNSs) which support several plasmon modes with a moderate size distribution. We use crystal violet~(CV) as the Raman molecule and CdSe/ZnS core-shell nanoparticles as the aux QDs. Our theoretical model successfully predicts the observed enhancements.

\section{Results}

AuNSs are solution synthesized following previously reported procedures \cite{emren_2018}. The Raman reporter molecule CV with/without the CdSe/ZnS QDs are in-solution mixed and dispersed on single side polished Si substrates. AuNSs are characterized using SEM/TEM for size determination and UV-Vis absorption spectrophotometry for extinction. The AuNSs are found to be 50 nm diameter on average and confirmed to possess a spiky star-shape as reported in literature. The UV-Vis spectrum displays a broad plasmonic response reminiscent of multiple contributing plasmon resonances as AuNS shape supports various similar length scales corresponding to closely neighboring peak centers at the same time (SM~\cite{supp} Fig. S2).

\begin{figure}
\includegraphics[width=0.48 \textwidth]{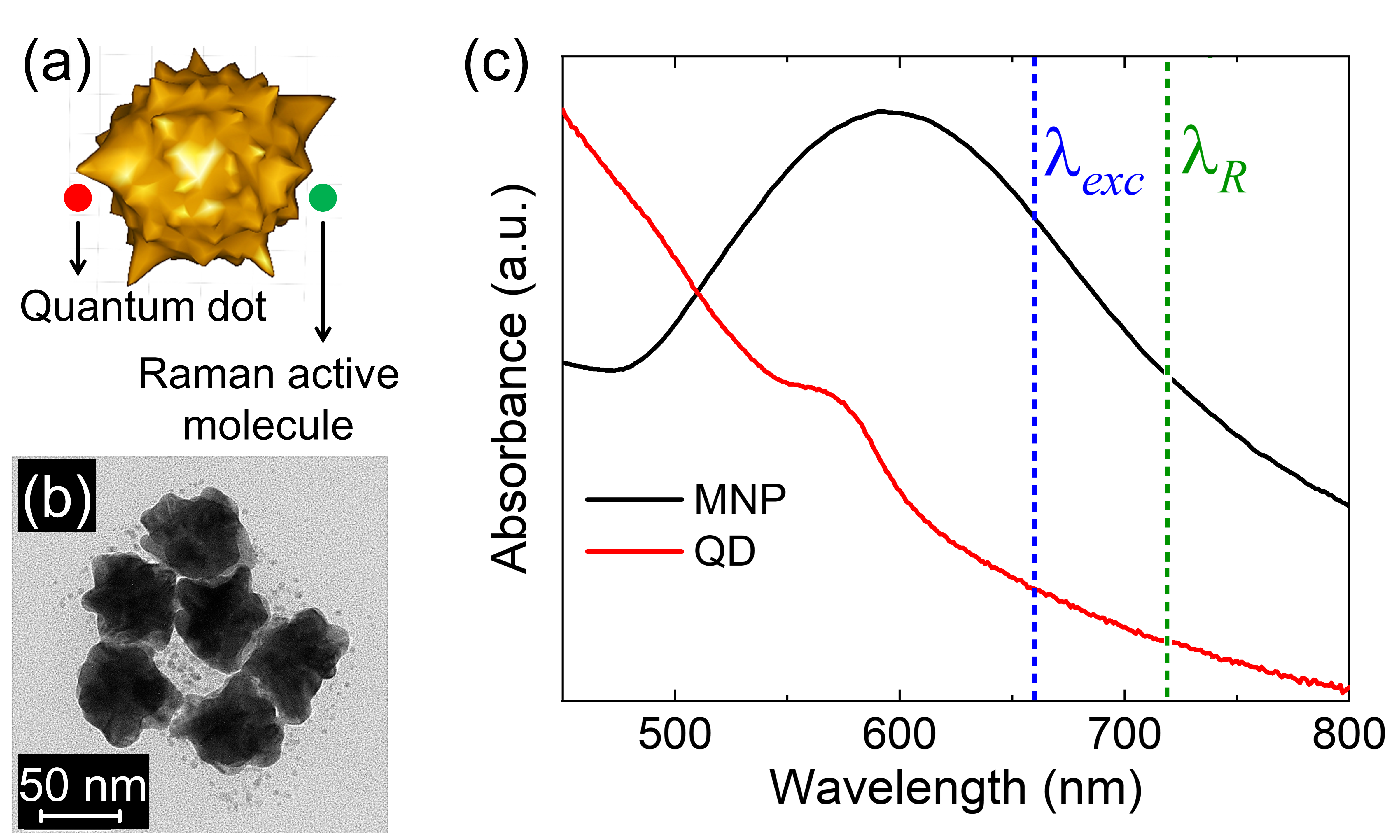}
\caption{\label{fig1} (a) A schematic reproduction of an AuNS. (b) TEM image of AuNSs.  (c) UV-Vis spectrum of QDs and AuNSs showing their absorption bands.}
\end{figure}

Figs. 1a and 1b show a close schematic representation of the AuNSs and a TEM image of a cluster of AuNSs used in the experiments, respectively. Fig. 1c shows the UV-Vis absorption spectra of the QDs (in red) and AuNSs (in black) used in the experiments. The excitation and Raman shifted wavelengths are indicated by vertical dashed blue ($\lambda_{\rm exc}$=660 nm) and green ($\lambda_{\scriptscriptstyle \rm R}$=715 nm) lines. The Raman shifted wavelength corresponds to the 1165 ${\rm cm}^{-1}$ peak used for demonstrating our results. 

Fig. 2a shows Raman spectra acquired in the presence of the AuNSs alone (in black), the AuNSs mixed with QDs (green), the AuNSs mixed with CV (blue), and AuNSs mixed with QDs and CV (red). The AuNSs and AuNS+QDs are shown to produce a merely flat baseline, where AuNSs are shown to generate a SERS enhancement and well-discernible CV spectrum is obtained; and the addition of QDs are shown to significantly further enhance the CV spectrum by almost an order of magnitude. We have mixed equal number of AuNSs and QDs in the quantum enhancement measurements in Fig. 2. All spectra are obtained under same operation conditions regarding the excitation power, focal spot size, accumulation time, etc. The raw data is only baseline subtracted for clarity without further processing. Note that even if the excitation wavelength of 660 nm does not fall within the absorption band of CV (430-630 nm), still a weak, continuous fluorescence background in the form of a linear baseline is observed in the Raman spectra. The flat Si substrate Raman peak at 521 ${\rm cm}^{-1}$ is indicated for reference. 

Fig. 2b provides a clear demonstration of the quantum~(path interference~\cite{postaci_silent_2018}) EFs normalized to the conventional SERS EFs. Two bar charts showing Raman intensities acquired at 10 different locations are provided for direct comparison of the two cases (without and with QDs). The figure displays an obvious and statistically significant difference in the mean Raman intensity separated by about a factor of $\sim$7 between the cases of without/with QDs which is the signature of the path interference effect, where other alternative explanations can be excluded (see the Conclusion section). The spectral range of 1120-1220 ${\rm cm}^{-1}$ is baseline corrected, curve fitted and area under the fit curve is used for each acquisition in Fig. 2b.

\begin{figure}
\includegraphics[width=0.48 \textwidth]{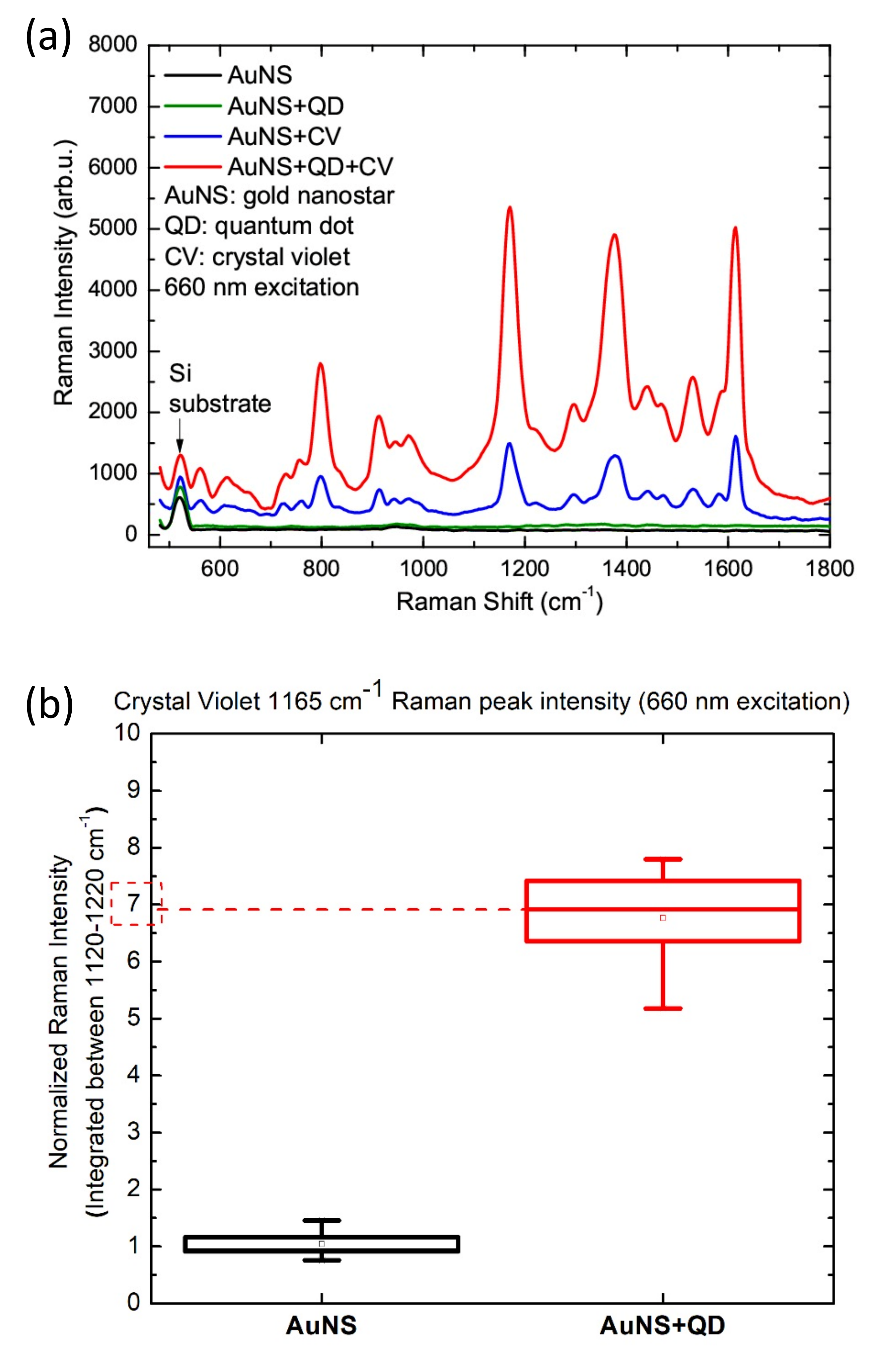}
\caption{\label{fig2} (a) Raman spectra of the AuNSs alone (black), the AuNSs mixed with QDs (green), the AuNSs mixed with CV (blue), and AuNSs mixed with QDs and CV (red), (b) Quantum EF normalized to conventional EF for AuNSs and AuNSs+QDs.}
\end{figure}

In Sec. I of the SM, we also briefly explain why additional~(QUPERS) EFs occur~\cite{postaci_silent_2018} on a simple equation (Eq.~(S1)). Furthermore, we captured the picture of a representative AuNS from the TEM images and perform 3D meshing. We determined the scattering/absorption cross-section of the AuNS using FDTD simulations. We performed peak fitting on its plasmonic response. See SM~\cite{supp} Fig. S2. We utilized the determined plasmon resonances to perform analytical simulations. We used various choices of the plasmon resonance doublets into which pump and Raman-shifted light couples. (See SM~\cite{supp} Table I.) We show that for a relatively broad interval of QD level-spacing values quantum enhancement manifests itself, see SM~\cite{supp} Fig. S1. Different choices of the doublets also lead to 1-2 orders-of-magnitude quantum EFs, see Table I in the SM. These EFs are calculated assuming that the QD resides optimally at the hotspot. Theory predicts weaker (but still existing) quantum EFs for smaller QD-AuNS couplings. See SM~\cite{supp} Fig. S1.

In addition, we repeated the quantum enhancement experiments using different ratios of QDs per AuNSs. The AuNS batch is a different one than that of the batch used in Fig. 2 with slightly different AuNS geometry in order to test the universality of our approach. In this configuration the maximum quantum EF is $\sim$3. Same CV Raman peak of 1165 ${\rm cm}^{-1}$ is used. Fig. 3a shows the experimentally determined average quantum EFs normalized to the conventional SERS EF, for different QD/AuNS ratios. We observe that the maximum quantum EF is attained at a ratio close to unity where it is weaker for lower or larger values. This is exactly as one would expect from the quantum enhancement effect where the EF is weaker when QDs are in absence or when more than one QD interacts on average with each AuNS. The second QD, attached to the AuNS at a different position in general, also introduces a path interference effect. But since the phases of the two QDs differ, the resulting effect is weaker due to destructive interference. Other CV Raman peaks are found to exhibit similar dependence also (See SM~\cite{supp} Fig S3). 

We provide a theoretical treatment of the cases in which more than one QD interacts with the same AuNS. It is observed that the phase differences in the path interferences weaken the quantum enhancement effect (SM, Sec. III). Different phases emerge due to retardation between the QDs residing at different relative locations on the AuNS. Furthermore, we also study the same effect using exact solutions of the Maxwell equations (FDTD) and in Fig. 3b we demonstrate that a similar pattern as in Fig. 3a is to be expected. For details see SM Sec. III. It should be noted that the treatment necessarily starts from one QD/AuNS and continues upwards with integer ratios, where the experimental values correspond to both average ratios and average EFs originating from arbitrary configurations.

\begin{figure}
\includegraphics[width=0.35 \textwidth]{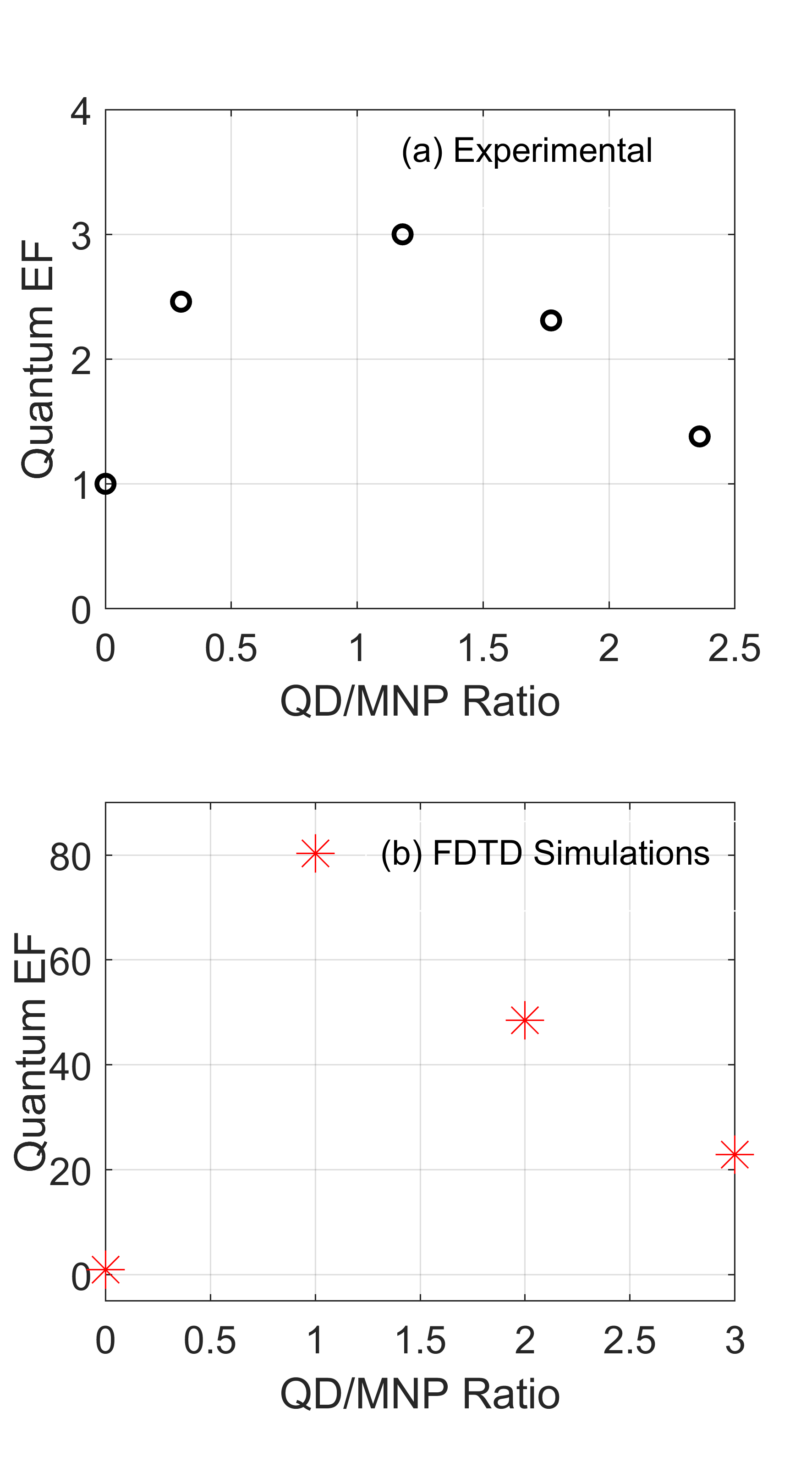}
\caption{(a) Experimentally determined quantum EFs for different ratios of QDs/AuNSs. (b) Numerically determined quantum EFs for different QD/AuNSs. Both experiment and theory display an optimal value centered around unity, a clear evidence for QUPERS, where the differing phases of QDs residing at different positions gives rise to weakening of the quantum enhancement effect due to destructive interference.}
\label{fig3}
\end{figure}

\section{Conclusions, Discussions and Outlook}

We present the first stochastic experimental demonstration of the path interference phenomenon taking place in a Raman process~\cite{postaci_silent_2018}. Presence of QDs multiplies the conventional (field enhancement-based) SERS signal on average 7 times~\cite{PS_3times_enhancement}. We show that extra EFs in our experiment can originate only from path (Fano) interference effect. Experimentally determined quantum enhancement, may appear small at first sight. It should be noted that the measured quantum EFs belong to a stochastic distribution of QDs over AuNSs where they do not necessarily reside at the hotspots. In single particle experiments where positions of QDs can be controlled using, e.g., complimentary chemistry, scanning probe techniques \cite{manipulating_QD}, or on-demand generation of QDs~\cite{hu2024quantum}, the quantum EFs are expected to reach 2-3 orders-of-magnitude~\cite{postaci_silent_2018}.

We present strong evidences (i)-(vi) demonstrating that the measured 7-fold quantum enhancement takes place solely due to path interference while excluding other mechanisms, leaving no space for doubt. (i) Absorption peak of the QDs ($\Omega_{\rm \scriptscriptstyle QD}$=562 nm) is off-resonant with the pump frequency $\omega$=660 nm, thus cannot promote the conversion directly. (ii) The strong fluorescence background that QDs may generate even with the off resonant 660 nm pump, is already subtracted. (iii) We also consider possibility of refractive index modulation due to the added QDs and show that it cannot result in an overall enhancement that can account for more than 5-10\%. (iv) Our theoretical description explains the spectral positions of the enhancements. (v) As the most significant evidence: SERS enhancement starts to decrease when the number of QDs per AuNS exceeds 1 as anticipated by the theory. This is because presence of a second (or more) QD on the same AuNS weakens the enhancement scheme. The phase introduced from the retardation of the second QD, with respect to the first QD, weakens the overall EF appearing due to the path interference effect. We demonstrate the existence of this phenomenon from both analytical and FDTD-based numerical simulations. (vi) Possibility of a chemical enhancement mechanism can be eliminated based on the same fact that the EF declines for QD/AuNS ratio larger than 1. This is not to be expected from polarizability tuning of CV and/or charge transfer, and also considering that there are exceedingly more CV molecules in the experimental setting than either the AuNSs or QDs. For a straightforward yet comprehensive comparison of the QUPERS effect with the normal Raman and SERS effects, we provide the three schemes in a sketch in SM~\cite{supp} Fig. S4.

It should be noted that the underlying mechanism in our demonstration differs itself from formerly reported Fano-based improved SERS systems~\cite{he2016near,zhang2014coherent,ye2012plasmonic} which are based on lifetime enhancement~\cite{Yildiz2020} via coupling to dark modes. Here, we study the enhancement of steady-state (CW pump) Raman amplitude via path interference effects that appear from coupling to a QD. (See SM Sec. IV for details.)
In order to benefit from the mechanism that we utilize for SERS, there is no requirement of using costly, time-consuming, complicated lithography based top-down approaches. Solution based low-cost, rapid yet stochastic bottom-up fabrication approach is sufficient. Moreover, we utilize continuous-wave excited steady state outcomes rather than ultra-fast excited, transient outcomes.

The phenomenon, we now also demonstrate experimentally, opens a new avenue in Raman based detection techniques. For instance, selective and actively-tunable enhancement of certain Raman bands as markers of a particular interest among a crowded spectrum acquired from a complex multi-component material matrix is one of the many possibilities. This can be achieved by controlled production of QDs near nanostructures ~\cite{hu2024quantum} combined with electrical-tuning of QD resonances~\cite{flatte2008giant,empedocles1997quantum,asif2024voltage,gunay2023demand}.

\begin{acknowledgments}

The authors acknowledge support from TUBITAK-1001 Project No. 119F101. MET gratefully thanks Fikri Isik.

\end{acknowledgments}

\bibliography{bibliography}

\begin{thebibliography}{30}%
\makeatletter
\providecommand \@ifxundefined [1]{%
 \@ifx{#1\undefined}
}%
\providecommand \@ifnum [1]{%
 \ifnum #1\expandafter \@firstoftwo
 \else \expandafter \@secondoftwo
 \fi
}%
\providecommand \@ifx [1]{%
 \ifx #1\expandafter \@firstoftwo
 \else \expandafter \@secondoftwo
 \fi
}%
\providecommand \natexlab [1]{#1}%
\providecommand \enquote  [1]{``#1''}%
\providecommand \bibnamefont  [1]{#1}%
\providecommand \bibfnamefont [1]{#1}%
\providecommand \citenamefont [1]{#1}%
\providecommand \href@noop [0]{\@secondoftwo}%
\providecommand \href [0]{\begingroup \@sanitize@url \@href}%
\providecommand \@href[1]{\@@startlink{#1}\@@href}%
\providecommand \@@href[1]{\endgroup#1\@@endlink}%
\providecommand \@sanitize@url [0]{\catcode `\\12\catcode `\$12\catcode
  `\&12\catcode `\#12\catcode `\^12\catcode `\_12\catcode `\%12\relax}%
\providecommand \@@startlink[1]{}%
\providecommand \@@endlink[0]{}%
\providecommand \url  [0]{\begingroup\@sanitize@url \@url }%
\providecommand \@url [1]{\endgroup\@href {#1}{\urlprefix }}%
\providecommand \urlprefix  [0]{URL }%
\providecommand \Eprint [0]{\href }%
\providecommand \doibase [0]{https://doi.org/}%
\providecommand \selectlanguage [0]{\@gobble}%
\providecommand \bibinfo  [0]{\@secondoftwo}%
\providecommand \bibfield  [0]{\@secondoftwo}%
\providecommand \translation [1]{[#1]}%
\providecommand \BibitemOpen [0]{}%
\providecommand \bibitemStop [0]{}%
\providecommand \bibitemNoStop [0]{.\EOS\space}%
\providecommand \EOS [0]{\spacefactor3000\relax}%
\providecommand \BibitemShut  [1]{\csname bibitem#1\endcsname}%
\let\auto@bib@innerbib\@empty
\bibitem [{\citenamefont {Betzig}\ and\ \citenamefont
  {Chichester}(1993)}]{snom_single_molecule}%
  \BibitemOpen
  \bibfield  {author} {\bibinfo {author} {\bibfnamefont {E.}~\bibnamefont
  {Betzig}}\ and\ \bibinfo {author} {\bibfnamefont {R.}~\bibnamefont
  {Chichester}},\ }\bibfield  {title} {\bibinfo {title} {Single molecules
  observed by near-field scanning optical microscopy},\ }\href@noop {}
  {\bibfield  {journal} {\bibinfo  {journal} {Science}\ }\textbf {\bibinfo
  {volume} {262}},\ \bibinfo {pages} {1422} (\bibinfo {year}
  {1993})}\BibitemShut {NoStop}%
\bibitem [{\citenamefont {Ovali}\ \emph {et~al.}(2021)\citenamefont {Ovali},
  \citenamefont {Sahin}, \citenamefont {Bek},\ and\ \citenamefont
  {Tasgin}}]{ovali_single_APL}%
  \BibitemOpen
  \bibfield  {author} {\bibinfo {author} {\bibfnamefont {R.~V.}\ \bibnamefont
  {Ovali}}, \bibinfo {author} {\bibfnamefont {R.}~\bibnamefont {Sahin}},
  \bibinfo {author} {\bibfnamefont {A.}~\bibnamefont {Bek}},\ and\ \bibinfo
  {author} {\bibfnamefont {M.~E.}\ \bibnamefont {Tasgin}},\ }\bibfield  {title}
  {\bibinfo {title} {Single-molecule-resolution ultrafast near-field optical
  microscopy via plasmon lifetime extension},\ }\href@noop {} {\bibfield
  {journal} {\bibinfo  {journal} {Applied Physics Letters}\ }\textbf {\bibinfo
  {volume} {118}} (\bibinfo {year} {2021})}\BibitemShut {NoStop}%
\bibitem [{\citenamefont {Grosse}\ \emph {et~al.}(2012)\citenamefont {Grosse},
  \citenamefont {Heckmann},\ and\ \citenamefont
  {Woggon}}]{grosse2012nonlinear}%
  \BibitemOpen
  \bibfield  {author} {\bibinfo {author} {\bibfnamefont {N.~B.}\ \bibnamefont
  {Grosse}}, \bibinfo {author} {\bibfnamefont {J.}~\bibnamefont {Heckmann}},\
  and\ \bibinfo {author} {\bibfnamefont {U.}~\bibnamefont {Woggon}},\
  }\bibfield  {title} {\bibinfo {title} {Nonlinear plasmon-photon interaction
  resolved by k-space spectroscopy},\ }\href@noop {} {\bibfield  {journal}
  {\bibinfo  {journal} {Physical Review Letters}\ }\textbf {\bibinfo {volume}
  {108}},\ \bibinfo {pages} {136802} (\bibinfo {year} {2012})}\BibitemShut
  {NoStop}%
\bibitem [{\citenamefont {Taşgın}\ \emph {et~al.}(2018)\citenamefont
  {Taşgın}, \citenamefont {Bek},\ and\ \citenamefont
  {Postacı}}]{bookchapter}%
  \BibitemOpen
  \bibfield  {author} {\bibinfo {author} {\bibfnamefont {M.~E.}\ \bibnamefont
  {Taşgın}}, \bibinfo {author} {\bibfnamefont {A.}~\bibnamefont {Bek}},\ and\
  \bibinfo {author} {\bibfnamefont {S.}~\bibnamefont {Postacı}},\ }\bibfield
  {title} {\bibinfo {title} {Fano resonances in the linear and nonlinear
  plasmonic response},\ }\href@noop {} {\bibfield  {journal} {\bibinfo
  {journal} {Springer Series in Optical Sciences}\ }\textbf {\bibinfo {volume}
  {219}},\ \bibinfo {pages} {1} (\bibinfo {year} {2018})}\BibitemShut {NoStop}%
\bibitem [{\citenamefont {Noor}\ \emph {et~al.}(2020)\citenamefont {Noor},
  \citenamefont {Damodaran}, \citenamefont {Lee}, \citenamefont {Maier},
  \citenamefont {Oh},\ and\ \citenamefont {Cirac{\`\i}}}]{noor2020mode}%
  \BibitemOpen
  \bibfield  {author} {\bibinfo {author} {\bibfnamefont {A.}~\bibnamefont
  {Noor}}, \bibinfo {author} {\bibfnamefont {A.~R.}\ \bibnamefont {Damodaran}},
  \bibinfo {author} {\bibfnamefont {I.-H.}\ \bibnamefont {Lee}}, \bibinfo
  {author} {\bibfnamefont {S.~A.}\ \bibnamefont {Maier}}, \bibinfo {author}
  {\bibfnamefont {S.-H.}\ \bibnamefont {Oh}},\ and\ \bibinfo {author}
  {\bibfnamefont {C.}~\bibnamefont {Cirac{\`\i}}},\ }\bibfield  {title}
  {\bibinfo {title} {Mode-matching enhancement of second-harmonic generation
  with plasmonic nanopatch antennas},\ }\href@noop {} {\bibfield  {journal}
  {\bibinfo  {journal} {ACS photonics}\ }\textbf {\bibinfo {volume} {7}},\
  \bibinfo {pages} {3333} (\bibinfo {year} {2020})}\BibitemShut {NoStop}%
\bibitem [{\citenamefont {Nie}\ and\ \citenamefont
  {Emory}(1997)}]{nie_single_molecule}%
  \BibitemOpen
  \bibfield  {author} {\bibinfo {author} {\bibfnamefont {S.}~\bibnamefont
  {Nie}}\ and\ \bibinfo {author} {\bibfnamefont {S.~R.}\ \bibnamefont
  {Emory}},\ }\bibfield  {title} {\bibinfo {title} {Probing single molecules
  and single nanoparticles by surface-enhanced raman scattering},\ }\href@noop
  {} {\bibfield  {journal} {\bibinfo  {journal} {Science}\ }\textbf {\bibinfo
  {volume} {275}},\ \bibinfo {pages} {1102} (\bibinfo {year}
  {1997})}\BibitemShut {NoStop}%
\bibitem [{\citenamefont {Postaci}\ \emph {et~al.}(2018)\citenamefont
  {Postaci}, \citenamefont {Yildiz}, \citenamefont {Bek},\ and\ \citenamefont
  {Tasgin}}]{postaci_silent_2018}%
  \BibitemOpen
  \bibfield  {author} {\bibinfo {author} {\bibfnamefont {S.}~\bibnamefont
  {Postaci}}, \bibinfo {author} {\bibfnamefont {B.~C.}\ \bibnamefont {Yildiz}},
  \bibinfo {author} {\bibfnamefont {A.}~\bibnamefont {Bek}},\ and\ \bibinfo
  {author} {\bibfnamefont {M.~E.}\ \bibnamefont {Tasgin}},\ }\bibfield  {title}
  {\bibinfo {title} {Silent enhancement of {SERS} signal without increasing hot
  spot intensities},\ }\href {https://doi.org/10.1515/nanoph-2018-0089}
  {\bibfield  {journal} {\bibinfo  {journal} {Nanophotonics}\ }\textbf
  {\bibinfo {volume} {7}},\ \bibinfo {pages} {1687} (\bibinfo {year}
  {2018})}\BibitemShut {NoStop}%
\bibitem [{\citenamefont {Gencaslan}\ \emph {et~al.}(2022)\citenamefont
  {Gencaslan}, \citenamefont {Aytas}, \citenamefont {Asif}, \citenamefont
  {Tasgin},\ and\ \citenamefont {Sahin}}]{gencaslan2022silent}%
  \BibitemOpen
  \bibfield  {author} {\bibinfo {author} {\bibfnamefont {A.}~\bibnamefont
  {Gencaslan}}, \bibinfo {author} {\bibfnamefont {T.~T.}\ \bibnamefont
  {Aytas}}, \bibinfo {author} {\bibfnamefont {H.}~\bibnamefont {Asif}},
  \bibinfo {author} {\bibfnamefont {M.~E.}\ \bibnamefont {Tasgin}},\ and\
  \bibinfo {author} {\bibfnamefont {R.}~\bibnamefont {Sahin}},\ }\bibfield
  {title} {\bibinfo {title} {Silent-enhancement of multiple raman modes via
  tuning optical properties of graphene nanostructures},\ }\href@noop {}
  {\bibfield  {journal} {\bibinfo  {journal} {The European Physical Journal
  Plus}\ }\textbf {\bibinfo {volume} {137}},\ \bibinfo {pages} {1} (\bibinfo
  {year} {2022})}\BibitemShut {NoStop}%
\bibitem [{\citenamefont {Schmidt}\ and\ \citenamefont
  {Imamoglu}(1996)}]{schmidt1996giant}%
  \BibitemOpen
  \bibfield  {author} {\bibinfo {author} {\bibfnamefont {H.}~\bibnamefont
  {Schmidt}}\ and\ \bibinfo {author} {\bibfnamefont {A.}~\bibnamefont
  {Imamoglu}},\ }\bibfield  {title} {\bibinfo {title} {Giant kerr
  nonlinearities obtained by electromagnetically induced transparency},\
  }\href@noop {} {\bibfield  {journal} {\bibinfo  {journal} {Optics letters}\
  }\textbf {\bibinfo {volume} {21}},\ \bibinfo {pages} {1936} (\bibinfo {year}
  {1996})}\BibitemShut {NoStop}%
\bibitem [{\citenamefont {Paspalakis}\ \emph {et~al.}(2014)\citenamefont
  {Paspalakis}, \citenamefont {Evangelou}, \citenamefont {Kosionis},\ and\
  \citenamefont {Terzis}}]{paspalakis2014strongly}%
  \BibitemOpen
  \bibfield  {author} {\bibinfo {author} {\bibfnamefont {E.}~\bibnamefont
  {Paspalakis}}, \bibinfo {author} {\bibfnamefont {S.}~\bibnamefont
  {Evangelou}}, \bibinfo {author} {\bibfnamefont {S.~G.}\ \bibnamefont
  {Kosionis}},\ and\ \bibinfo {author} {\bibfnamefont {A.~F.}\ \bibnamefont
  {Terzis}},\ }\bibfield  {title} {\bibinfo {title} {Strongly modified
  four-wave mixing in a coupled semiconductor quantum dot-metal nanoparticle
  system},\ }\href@noop {} {\bibfield  {journal} {\bibinfo  {journal} {Journal
  of Applied Physics}\ }\textbf {\bibinfo {volume} {115}} (\bibinfo {year}
  {2014})}\BibitemShut {NoStop}%
\bibitem [{\citenamefont {Singh}\ \emph {et~al.}(2016)\citenamefont {Singh},
  \citenamefont {Abak},\ and\ \citenamefont {Tasgin}}]{singh2016enhancement}%
  \BibitemOpen
  \bibfield  {author} {\bibinfo {author} {\bibfnamefont {S.}~\bibnamefont
  {Singh}}, \bibinfo {author} {\bibfnamefont {M.}~\bibnamefont {Abak}},\ and\
  \bibinfo {author} {\bibfnamefont {M.}~\bibnamefont {Tasgin}},\ }\bibfield
  {title} {\bibinfo {title} {Enhancement of four-wave mixing via interference
  of multiple plasmonic conversion paths},\ }\href@noop {} {\bibfield
  {journal} {\bibinfo  {journal} {Physical Review B}\ }\textbf {\bibinfo
  {volume} {93}},\ \bibinfo {pages} {035410} (\bibinfo {year}
  {2016})}\BibitemShut {NoStop}%
\bibitem [{\citenamefont {Scully}\ and\ \citenamefont
  {Zubairy}(1997)}]{ScullyZubairyBook}%
  \BibitemOpen
  \bibfield  {author} {\bibinfo {author} {\bibfnamefont {M.~O.}\ \bibnamefont
  {Scully}}\ and\ \bibinfo {author} {\bibfnamefont {M.~S.}\ \bibnamefont
  {Zubairy}},\ }\href
  {https://www.cambridge.org/core/books/quantum-optics/08DC53888452CBC6CDC0FD8A1A1A4DD7}
  {\emph {\bibinfo {title} {Quantum Optics}}}\ (\bibinfo  {publisher}
  {Cambridge University Press},\ \bibinfo {address} {New York},\ \bibinfo
  {year} {1997})\BibitemShut {NoStop}%
\bibitem [{PS_({\natexlab{a}})}]{PS_plasmon_tuning}%
  \BibitemOpen
  \href@noop {} {} ({\natexlab{a}}),\ \bibinfo {note} {{Aside from
  electrochromic, thermoelectric, etc tuning of dielectric function around the
  NP, VO5 \cite{li_2019} which effects entire plasmon mode
  spectrum.}}\BibitemShut {Stop}%
\bibitem [{PS_({\natexlab{b}})}]{PS_nanostars}%
  \BibitemOpen
  \href@noop {} {} ({\natexlab{b}}),\ \bibinfo {note} {{In our experimental
  setting we have employed nanostars that supports a multitude of plasmon modes
  (see Fig.~S2 in SM) in order to ensure to hit two resonant plasmon modes,
  because the work is the first of its kind and proof-of-principle is the
  priority. This however entails our setup coverage of many plasmon modes at
  the same time obscuring the experimental demonstration of mode
  selectivity.}}\BibitemShut {Stop}%
\bibitem [{\citenamefont {Paresh Chandra~RAY}\ and\ \citenamefont
  {FU}(2011)}]{paresh_2011}%
  \BibitemOpen
  \bibfield  {author} {\bibinfo {author} {\bibfnamefont {H.~Y.}\ \bibnamefont
  {Paresh Chandra~RAY}}\ and\ \bibinfo {author} {\bibfnamefont {P.~P.}\
  \bibnamefont {FU}},\ }\bibfield  {title} {\bibinfo {title} {Nanogold-based
  sensing of environmental toxins: Excitement and challenges},\ }\href@noop {}
  {\bibfield  {journal} {\bibinfo  {journal} {Journal of Environmental Science
  and Health, Part C}\ }\textbf {\bibinfo {volume} {29}},\ \bibinfo {pages}
  {52} (\bibinfo {year} {2011})}\BibitemShut {NoStop}%
\bibitem [{\citenamefont {Numan}\ \emph {et~al.}(2022)\citenamefont {Numan},
  \citenamefont {Singh}, \citenamefont {Alam}, \citenamefont {Khalid},
  \citenamefont {Li},\ and\ \citenamefont {Singh}}]{numan_2022}%
  \BibitemOpen
  \bibfield  {author} {\bibinfo {author} {\bibfnamefont {A.}~\bibnamefont
  {Numan}}, \bibinfo {author} {\bibfnamefont {P.~S.}\ \bibnamefont {Singh}},
  \bibinfo {author} {\bibfnamefont {A.}~\bibnamefont {Alam}}, \bibinfo {author}
  {\bibfnamefont {M.}~\bibnamefont {Khalid}}, \bibinfo {author} {\bibfnamefont
  {L.}~\bibnamefont {Li}},\ and\ \bibinfo {author} {\bibfnamefont
  {S.}~\bibnamefont {Singh}},\ }\bibfield  {title} {\bibinfo {title} {Advances
  in noble-metal nanoparticle-based fluorescence detection of organophosphorus
  chemical warfare agents},\ }\href@noop {} {\bibfield  {journal} {\bibinfo
  {journal} {ACS Omega}\ }\textbf {\bibinfo {volume} {7}},\ \bibinfo {pages}
  {27079} (\bibinfo {year} {2022})}\BibitemShut {NoStop}%
\bibitem [{\citenamefont {Flatt{\'e}}\ \emph {et~al.}(2008)\citenamefont
  {Flatt{\'e}}, \citenamefont {Kornyshev},\ and\ \citenamefont
  {Urbakh}}]{flatte2008giant}%
  \BibitemOpen
  \bibfield  {author} {\bibinfo {author} {\bibfnamefont {M.}~\bibnamefont
  {Flatt{\'e}}}, \bibinfo {author} {\bibfnamefont {A.}~\bibnamefont
  {Kornyshev}},\ and\ \bibinfo {author} {\bibfnamefont {M.}~\bibnamefont
  {Urbakh}},\ }\bibfield  {title} {\bibinfo {title} {Giant stark effect in
  quantum dots at liquid/liquid interfaces: A new option for tunable optical
  filters},\ }\href@noop {} {\bibfield  {journal} {\bibinfo  {journal}
  {Proceedings of the National Academy of Sciences}\ }\textbf {\bibinfo
  {volume} {105}},\ \bibinfo {pages} {18212} (\bibinfo {year}
  {2008})}\BibitemShut {NoStop}%
\bibitem [{\citenamefont {Empedocles}\ and\ \citenamefont
  {Bawendi}(1997)}]{empedocles1997quantum}%
  \BibitemOpen
  \bibfield  {author} {\bibinfo {author} {\bibfnamefont {S.~A.}\ \bibnamefont
  {Empedocles}}\ and\ \bibinfo {author} {\bibfnamefont {M.~G.}\ \bibnamefont
  {Bawendi}},\ }\bibfield  {title} {\bibinfo {title} {Quantum-confined stark
  effect in single cdse nanocrystallite quantum dots},\ }\href@noop {}
  {\bibfield  {journal} {\bibinfo  {journal} {Science}\ }\textbf {\bibinfo
  {volume} {278}},\ \bibinfo {pages} {2114} (\bibinfo {year}
  {1997})}\BibitemShut {NoStop}%
\bibitem [{\citenamefont {Asif}\ \emph {et~al.}(2024)\citenamefont {Asif},
  \citenamefont {Bek}, \citenamefont {Tasgin},\ and\ \citenamefont
  {Sahin}}]{asif2024voltage}%
  \BibitemOpen
  \bibfield  {author} {\bibinfo {author} {\bibfnamefont {H.}~\bibnamefont
  {Asif}}, \bibinfo {author} {\bibfnamefont {A.}~\bibnamefont {Bek}}, \bibinfo
  {author} {\bibfnamefont {M.~E.}\ \bibnamefont {Tasgin}},\ and\ \bibinfo
  {author} {\bibfnamefont {R.}~\bibnamefont {Sahin}},\ }\bibfield  {title}
  {\bibinfo {title} {Voltage-controlled extraordinary optical transmission in
  the visible regime},\ }\href@noop {} {\bibfield  {journal} {\bibinfo
  {journal} {Physical Review B}\ }\textbf {\bibinfo {volume} {109}},\ \bibinfo
  {pages} {125425} (\bibinfo {year} {2024})}\BibitemShut {NoStop}%
\bibitem [{\citenamefont {G{\"u}nay}\ \emph {et~al.}(2023)\citenamefont
  {G{\"u}nay}, \citenamefont {Das}, \citenamefont {Y{\"u}ce}, \citenamefont
  {Polat}, \citenamefont {Bek},\ and\ \citenamefont
  {Tasgin}}]{gunay2023demand}%
  \BibitemOpen
  \bibfield  {author} {\bibinfo {author} {\bibfnamefont {M.}~\bibnamefont
  {G{\"u}nay}}, \bibinfo {author} {\bibfnamefont {P.}~\bibnamefont {Das}},
  \bibinfo {author} {\bibfnamefont {E.}~\bibnamefont {Y{\"u}ce}}, \bibinfo
  {author} {\bibfnamefont {E.~O.}\ \bibnamefont {Polat}}, \bibinfo {author}
  {\bibfnamefont {A.}~\bibnamefont {Bek}},\ and\ \bibinfo {author}
  {\bibfnamefont {M.~E.}\ \bibnamefont {Tasgin}},\ }\bibfield  {title}
  {\bibinfo {title} {On-demand continuous-variable quantum entanglement source
  for integrated circuits},\ }\href@noop {} {\bibfield  {journal} {\bibinfo
  {journal} {Nanophotonics}\ }\textbf {\bibinfo {volume} {12}},\ \bibinfo
  {pages} {229} (\bibinfo {year} {2023})}\BibitemShut {NoStop}%
\bibitem [{PS_({\natexlab{c}})}]{PS_3times_enhancement}%
  \BibitemOpen
  \href@noop {} {} ({\natexlab{c}}),\ \bibinfo {note} {{The actual Fano
  enhancement factors ---multiplying the usual SERS enhancement--- are much
  higher, e.g., about 2 orders of magnitude. However, in our experiment, QDs
  are distributed stochastically on nanostars. Most of them either do not
  introduce Fano interference due to a smaller coupling strength~(not hit
  hostpot) or introduces only a weak one. }}\BibitemShut {NoStop}%
\bibitem [{\citenamefont {Elci}\ \emph {et~al.}(2018)\citenamefont {Elci},
  \citenamefont {Demirtas}, \citenamefont {Ozturk}, \citenamefont {Bek},\ and\
  \citenamefont {Esenturk}}]{emren_2018}%
  \BibitemOpen
  \bibfield  {author} {\bibinfo {author} {\bibfnamefont {A.}~\bibnamefont
  {Elci}}, \bibinfo {author} {\bibfnamefont {O.}~\bibnamefont {Demirtas}},
  \bibinfo {author} {\bibfnamefont {I.~M.}\ \bibnamefont {Ozturk}}, \bibinfo
  {author} {\bibfnamefont {A.}~\bibnamefont {Bek}},\ and\ \bibinfo {author}
  {\bibfnamefont {E.~N.}\ \bibnamefont {Esenturk}},\ }\bibfield  {title}
  {\bibinfo {title} {Synthesis of tin oxide-coated gold nanostars and
  evaluation of their surface-enhanced raman scattering activities},\
  }\href@noop {} {\bibfield  {journal} {\bibinfo  {journal} {Journal of
  Materials Science}\ }\textbf {\bibinfo {volume} {53}},\ \bibinfo {pages}
  {16345} (\bibinfo {year} {2018})}\BibitemShut {NoStop}%
\bibitem [{sup()}]{supp}%
  \BibitemOpen
  \href@noop {} {}\bibinfo {note} {See Supplemental Material, link}\BibitemShut
  {NoStop}%
\bibitem [{\citenamefont {Ratchford}\ \emph {et~al.}(2011)\citenamefont
  {Ratchford}, \citenamefont {Shafiei}, \citenamefont {Kim}, \citenamefont
  {Gray},\ and\ \citenamefont {Li}}]{manipulating_QD}%
  \BibitemOpen
  \bibfield  {author} {\bibinfo {author} {\bibfnamefont {D.}~\bibnamefont
  {Ratchford}}, \bibinfo {author} {\bibfnamefont {F.}~\bibnamefont {Shafiei}},
  \bibinfo {author} {\bibfnamefont {S.}~\bibnamefont {Kim}}, \bibinfo {author}
  {\bibfnamefont {S.~K.}\ \bibnamefont {Gray}},\ and\ \bibinfo {author}
  {\bibfnamefont {X.}~\bibnamefont {Li}},\ }\bibfield  {title} {\bibinfo
  {title} {Manipulating coupling between a single semiconductor quantum dot and
  single gold nanoparticle},\ }\href@noop {} {\bibfield  {journal} {\bibinfo
  {journal} {Nano Letters}\ }\textbf {\bibinfo {volume} {11}},\ \bibinfo
  {pages} {1049} (\bibinfo {year} {2011})}\BibitemShut {NoStop}%
\bibitem [{\citenamefont {Hu}\ \emph {et~al.}(2024)\citenamefont {Hu},
  \citenamefont {Lorchat}, \citenamefont {Chen}, \citenamefont {Watanabe},
  \citenamefont {Taniguchi}, \citenamefont {Heinz}, \citenamefont {Murthy},\
  and\ \citenamefont {Chervy}}]{hu2024quantum}%
  \BibitemOpen
  \bibfield  {author} {\bibinfo {author} {\bibfnamefont {J.}~\bibnamefont
  {Hu}}, \bibinfo {author} {\bibfnamefont {E.}~\bibnamefont {Lorchat}},
  \bibinfo {author} {\bibfnamefont {X.}~\bibnamefont {Chen}}, \bibinfo {author}
  {\bibfnamefont {K.}~\bibnamefont {Watanabe}}, \bibinfo {author}
  {\bibfnamefont {T.}~\bibnamefont {Taniguchi}}, \bibinfo {author}
  {\bibfnamefont {T.~F.}\ \bibnamefont {Heinz}}, \bibinfo {author}
  {\bibfnamefont {P.~A.}\ \bibnamefont {Murthy}},\ and\ \bibinfo {author}
  {\bibfnamefont {T.}~\bibnamefont {Chervy}},\ }\bibfield  {title} {\bibinfo
  {title} {Quantum control of exciton wave functions in 2d semiconductors},\
  }\href@noop {} {\bibfield  {journal} {\bibinfo  {journal} {Science Advances}\
  }\textbf {\bibinfo {volume} {10}},\ \bibinfo {pages} {eadk6369} (\bibinfo
  {year} {2024})}\BibitemShut {NoStop}%
\bibitem [{\citenamefont {He}\ \emph {et~al.}(2016)\citenamefont {He},
  \citenamefont {Fan}, \citenamefont {Ding}, \citenamefont {Zhu},\ and\
  \citenamefont {Liang}}]{he2016near}%
  \BibitemOpen
  \bibfield  {author} {\bibinfo {author} {\bibfnamefont {J.}~\bibnamefont
  {He}}, \bibinfo {author} {\bibfnamefont {C.}~\bibnamefont {Fan}}, \bibinfo
  {author} {\bibfnamefont {P.}~\bibnamefont {Ding}}, \bibinfo {author}
  {\bibfnamefont {S.}~\bibnamefont {Zhu}},\ and\ \bibinfo {author}
  {\bibfnamefont {E.}~\bibnamefont {Liang}},\ }\bibfield  {title} {\bibinfo
  {title} {Near-field engineering of fano resonances in a plasmonic assembly
  for maximizing cars enhancements},\ }\href@noop {} {\bibfield  {journal}
  {\bibinfo  {journal} {Scientific reports}\ }\textbf {\bibinfo {volume} {6}},\
  \bibinfo {pages} {20777} (\bibinfo {year} {2016})}\BibitemShut {NoStop}%
\bibitem [{\citenamefont {Zhang}\ \emph {et~al.}(2014)\citenamefont {Zhang},
  \citenamefont {Zhen}, \citenamefont {Neumann}, \citenamefont {Day},
  \citenamefont {Nordlander},\ and\ \citenamefont {Halas}}]{zhang2014coherent}%
  \BibitemOpen
  \bibfield  {author} {\bibinfo {author} {\bibfnamefont {Y.}~\bibnamefont
  {Zhang}}, \bibinfo {author} {\bibfnamefont {Y.-R.}\ \bibnamefont {Zhen}},
  \bibinfo {author} {\bibfnamefont {O.}~\bibnamefont {Neumann}}, \bibinfo
  {author} {\bibfnamefont {J.~K.}\ \bibnamefont {Day}}, \bibinfo {author}
  {\bibfnamefont {P.}~\bibnamefont {Nordlander}},\ and\ \bibinfo {author}
  {\bibfnamefont {N.~J.}\ \bibnamefont {Halas}},\ }\bibfield  {title} {\bibinfo
  {title} {Coherent anti-stokes raman scattering with single-molecule
  sensitivity using a plasmonic fano resonance},\ }\href@noop {} {\bibfield
  {journal} {\bibinfo  {journal} {Nature communications}\ }\textbf {\bibinfo
  {volume} {5}},\ \bibinfo {pages} {1} (\bibinfo {year} {2014})}\BibitemShut
  {NoStop}%
\bibitem [{\citenamefont {Ye}\ \emph {et~al.}(2012)\citenamefont {Ye},
  \citenamefont {Wen}, \citenamefont {Sobhani}, \citenamefont {Lassiter},
  \citenamefont {Van~Dorpe}, \citenamefont {Nordlander},\ and\ \citenamefont
  {Halas}}]{ye2012plasmonic}%
  \BibitemOpen
  \bibfield  {author} {\bibinfo {author} {\bibfnamefont {J.}~\bibnamefont
  {Ye}}, \bibinfo {author} {\bibfnamefont {F.}~\bibnamefont {Wen}}, \bibinfo
  {author} {\bibfnamefont {H.}~\bibnamefont {Sobhani}}, \bibinfo {author}
  {\bibfnamefont {J.~B.}\ \bibnamefont {Lassiter}}, \bibinfo {author}
  {\bibfnamefont {P.}~\bibnamefont {Van~Dorpe}}, \bibinfo {author}
  {\bibfnamefont {P.}~\bibnamefont {Nordlander}},\ and\ \bibinfo {author}
  {\bibfnamefont {N.~J.}\ \bibnamefont {Halas}},\ }\bibfield  {title} {\bibinfo
  {title} {Plasmonic nanoclusters: near field properties of the fano resonance
  interrogated with sers},\ }\href@noop {} {\bibfield  {journal} {\bibinfo
  {journal} {Nano Letters}\ }\textbf {\bibinfo {volume} {12}},\ \bibinfo
  {pages} {1660} (\bibinfo {year} {2012})}\BibitemShut {NoStop}%
\bibitem [{\citenamefont {Yildiz}\ \emph {et~al.}(2020)\citenamefont {Yildiz},
  \citenamefont {Bek},\ and\ \citenamefont {Tasgin}}]{Yildiz2020}%
  \BibitemOpen
  \bibfield  {author} {\bibinfo {author} {\bibfnamefont {B.~C.}\ \bibnamefont
  {Yildiz}}, \bibinfo {author} {\bibfnamefont {A.}~\bibnamefont {Bek}},\ and\
  \bibinfo {author} {\bibfnamefont {M.~E.}\ \bibnamefont {Tasgin}},\ }\bibfield
   {title} {\bibinfo {title} {Plasmon lifetime enhancement in a bright-dark
  mode coupled system},\ }\href@noop {} {\bibfield  {journal} {\bibinfo
  {journal} {Physical Review B}\ }\textbf {\bibinfo {volume} {101}},\ \bibinfo
  {pages} {035416} (\bibinfo {year} {2020})}\BibitemShut {NoStop}%
\bibitem [{\citenamefont {Li}\ \emph {et~al.}(2019)\citenamefont {Li},
  \citenamefont {van~de Groep}, \citenamefont {Talin},\ and\ \citenamefont
  {Brongersma}}]{li_2019}%
  \BibitemOpen
  \bibfield  {author} {\bibinfo {author} {\bibfnamefont {Y.}~\bibnamefont
  {Li}}, \bibinfo {author} {\bibfnamefont {J.}~\bibnamefont {van~de Groep}},
  \bibinfo {author} {\bibfnamefont {A.~A.}\ \bibnamefont {Talin}},\ and\
  \bibinfo {author} {\bibfnamefont {M.~L.}\ \bibnamefont {Brongersma}},\
  }\bibfield  {title} {\bibinfo {title} {Dynamic tuning of gap plasmon
  resonances using a solid-state electrochromic device},\ }\href@noop {}
  {\bibfield  {journal} {\bibinfo  {journal} {Nano Letters}\ }\textbf {\bibinfo
  {volume} {19}},\ \bibinfo {pages} {7988} (\bibinfo {year}
  {2019})}\BibitemShut {NoStop}%
\end{thebibliography}%


\begin{thebibliography}{15}%
\makeatletter
\providecommand \@ifxundefined [1]{%
 \@ifx{#1\undefined}
}%
\providecommand \@ifnum [1]{%
 \ifnum #1\expandafter \@firstoftwo
 \else \expandafter \@secondoftwo
 \fi
}%
\providecommand \@ifx [1]{%
 \ifx #1\expandafter \@firstoftwo
 \else \expandafter \@secondoftwo
 \fi
}%
\providecommand \natexlab [1]{#1}%
\providecommand \enquote  [1]{``#1''}%
\providecommand \bibnamefont  [1]{#1}%
\providecommand \bibfnamefont [1]{#1}%
\providecommand \citenamefont [1]{#1}%
\providecommand \href@noop [0]{\@secondoftwo}%
\providecommand \href [0]{\begingroup \@sanitize@url \@href}%
\providecommand \@href[1]{\@@startlink{#1}\@@href}%
\providecommand \@@href[1]{\endgroup#1\@@endlink}%
\providecommand \@sanitize@url [0]{\catcode `\\12\catcode `\$12\catcode
  `\&12\catcode `\#12\catcode `\^12\catcode `\_12\catcode `\%12\relax}%
\providecommand \@@startlink[1]{}%
\providecommand \@@endlink[0]{}%
\providecommand \url  [0]{\begingroup\@sanitize@url \@url }%
\providecommand \@url [1]{\endgroup\@href {#1}{\urlprefix }}%
\providecommand \urlprefix  [0]{URL }%
\providecommand \Eprint [0]{\href }%
\providecommand \doibase [0]{https://doi.org/}%
\providecommand \selectlanguage [0]{\@gobble}%
\providecommand \bibinfo  [0]{\@secondoftwo}%
\providecommand \bibfield  [0]{\@secondoftwo}%
\providecommand \translation [1]{[#1]}%
\providecommand \BibitemOpen [0]{}%
\providecommand \bibitemStop [0]{}%
\providecommand \bibitemNoStop [0]{.\EOS\space}%
\providecommand \EOS [0]{\spacefactor3000\relax}%
\providecommand \BibitemShut  [1]{\csname bibitem#1\endcsname}%
\let\auto@bib@innerbib\@empty
\bibitem [{\citenamefont {Postaci}\ \emph {et~al.}(2018)\citenamefont
  {Postaci}, \citenamefont {Yildiz}, \citenamefont {Bek},\ and\ \citenamefont
  {Tasgin}}]{postaci_silent_2018}%
  \BibitemOpen
  \bibfield  {author} {\bibinfo {author} {\bibfnamefont {S.}~\bibnamefont
  {Postaci}}, \bibinfo {author} {\bibfnamefont {B.~C.}\ \bibnamefont {Yildiz}},
  \bibinfo {author} {\bibfnamefont {A.}~\bibnamefont {Bek}},\ and\ \bibinfo
  {author} {\bibfnamefont {M.~E.}\ \bibnamefont {Tasgin}},\ }\bibfield  {title}
  {\bibinfo {title} {Silent enhancement of {SERS} signal without increasing hot
  spot intensities},\ }\href {https://doi.org/10.1515/nanoph-2018-0089}
  {\bibfield  {journal} {\bibinfo  {journal} {Nanophotonics}\ }\textbf
  {\bibinfo {volume} {7}},\ \bibinfo {pages} {1687} (\bibinfo {year}
  {2018})}\BibitemShut {NoStop}%
\bibitem [{\citenamefont {Grosse}\ \emph {et~al.}(2012)\citenamefont {Grosse},
  \citenamefont {Heckmann},\ and\ \citenamefont
  {Woggon}}]{grosse2012nonlinear}%
  \BibitemOpen
  \bibfield  {author} {\bibinfo {author} {\bibfnamefont {N.~B.}\ \bibnamefont
  {Grosse}}, \bibinfo {author} {\bibfnamefont {J.}~\bibnamefont {Heckmann}},\
  and\ \bibinfo {author} {\bibfnamefont {U.}~\bibnamefont {Woggon}},\
  }\bibfield  {title} {\bibinfo {title} {Nonlinear plasmon-photon interaction
  resolved by $k$-space spectroscopy},\ }\href
  {https://doi.org/10.1103/PhysRevLett.108.136802} {\bibfield  {journal}
  {\bibinfo  {journal} {Phys. Rev. Lett.}\ }\textbf {\bibinfo {volume} {108}},\
  \bibinfo {pages} {136802} (\bibinfo {year} {2012})}\BibitemShut {NoStop}%
\bibitem [{\citenamefont {Taşgın}\ \emph {et~al.}(2018)\citenamefont
  {Taşgın}, \citenamefont {Bek},\ and\ \citenamefont
  {Postacı}}]{bookchapter}%
  \BibitemOpen
  \bibfield  {author} {\bibinfo {author} {\bibfnamefont {M.~E.}\ \bibnamefont
  {Taşgın}}, \bibinfo {author} {\bibfnamefont {A.}~\bibnamefont {Bek}},\ and\
  \bibinfo {author} {\bibfnamefont {S.}~\bibnamefont {Postacı}},\ }\bibfield
  {title} {\bibinfo {title} {Fano resonances in the linear and nonlinear
  plasmonic response},\ }\href@noop {} {\bibfield  {journal} {\bibinfo
  {journal} {Springer Series in Optical Sciences}\ }\textbf {\bibinfo {volume}
  {219}},\ \bibinfo {pages} {1} (\bibinfo {year} {2018})}\BibitemShut {NoStop}%
\bibitem [{\citenamefont {Kauranen}\ and\ \citenamefont
  {Zayats}(2012)}]{nonLinear_plasmonics_Kauranen2012}%
  \BibitemOpen
  \bibfield  {author} {\bibinfo {author} {\bibfnamefont {M.}~\bibnamefont
  {Kauranen}}\ and\ \bibinfo {author} {\bibfnamefont {A.~V.}\ \bibnamefont
  {Zayats}},\ }\bibfield  {title} {\bibinfo {title} {{Nonlinear plasmonics}},\
  }\href@noop {} {\bibfield  {journal} {\bibinfo  {journal} {Nature Photonics}\
  }\textbf {\bibinfo {volume} {6}},\ \bibinfo {pages} {737} (\bibinfo {year}
  {2012})}\BibitemShut {NoStop}%
\bibitem [{\citenamefont {Premaratne}\ and\ \citenamefont
  {Stockman}(2017)}]{Premaratne2017}%
  \BibitemOpen
  \bibfield  {author} {\bibinfo {author} {\bibfnamefont {M.}~\bibnamefont
  {Premaratne}}\ and\ \bibinfo {author} {\bibfnamefont {M.}~\bibnamefont
  {Stockman}},\ }\href {https://doi.org/10.1364/AOP.9.000079} {\emph {\bibinfo
  {title} {Adv. Opt. Photon.}}},\ Vol.~\bibinfo {volume} {9}\ (\bibinfo {year}
  {2017})\ pp.\ \bibinfo {pages} {79--128}\BibitemShut {NoStop}%
\bibitem [{\citenamefont {Hohenester}\ and\ \citenamefont
  {Trügler}(2012)}]{hohonester_MNPBEM}%
  \BibitemOpen
  \bibfield  {author} {\bibinfo {author} {\bibfnamefont {U.}~\bibnamefont
  {Hohenester}}\ and\ \bibinfo {author} {\bibfnamefont {A.}~\bibnamefont
  {Trügler}},\ }\bibfield  {title} {\bibinfo {title} {{M}{N}{P}{B}{E}{M} – a
  matlab toolbox for the simulation of plasmonic nanoparticles},\ }\href
  {https://doi.org/https://doi.org/10.1016/j.cpc.2011.09.009} {\bibfield
  {journal} {\bibinfo  {journal} {Computer Physics Communications}\ }\textbf
  {\bibinfo {volume} {183}},\ \bibinfo {pages} {370} (\bibinfo {year}
  {2012})}\BibitemShut {NoStop}%
\bibitem [{\citenamefont {Johnson}\ and\ \citenamefont
  {Christy}(1972)}]{au_dielectric}%
  \BibitemOpen
  \bibfield  {author} {\bibinfo {author} {\bibfnamefont {P.~B.}\ \bibnamefont
  {Johnson}}\ and\ \bibinfo {author} {\bibfnamefont {R.~W.}\ \bibnamefont
  {Christy}},\ }\bibfield  {title} {\bibinfo {title} {Optical constants of the
  noble metals},\ }\href@noop {} {\bibfield  {journal} {\bibinfo  {journal}
  {Phys. Rev. B}\ }\textbf {\bibinfo {volume} {6}},\ \bibinfo {pages} {4370}
  (\bibinfo {year} {1972})}\BibitemShut {NoStop}%
\bibitem [{\citenamefont {Tasgin}\ \emph {et~al.}(2018)\citenamefont {Tasgin},
  \citenamefont {Bek},\ and\ \citenamefont {Postaci}}]{tasginlinear2018}%
  \BibitemOpen
  \bibfield  {author} {\bibinfo {author} {\bibfnamefont {M.}~\bibnamefont
  {Tasgin}}, \bibinfo {author} {\bibfnamefont {A.}~\bibnamefont {Bek}},\ and\
  \bibinfo {author} {\bibfnamefont {S.}~\bibnamefont {Postaci}},\ }\href@noop
  {} {\emph {\bibinfo {title} {{F}ano Resonances in Optics and Microwaves:
  Physics and Application}}}\ (\bibinfo  {publisher} {Springer Review Book},\
  \bibinfo {year} {dated to be published in 2018})\ \bibinfo {note} {{{B}ook
  {C}hapter 1: ``{F}ano Resonances in the Linear and Nonlinear Plasmonic
  Response''}}\BibitemShut {NoStop}%
\bibitem [{\citenamefont {Hohenester}\ and\ \citenamefont
  {Tr{\"u}gler}(2012)}]{hohenester2012mnpbem}%
  \BibitemOpen
  \bibfield  {author} {\bibinfo {author} {\bibfnamefont {U.}~\bibnamefont
  {Hohenester}}\ and\ \bibinfo {author} {\bibfnamefont {A.}~\bibnamefont
  {Tr{\"u}gler}},\ }\bibfield  {title} {\bibinfo {title} {{M}{N}{P}{B}{E}{M}--a
  matlab toolbox for the simulation of plasmonic nanoparticles},\ }\href@noop
  {} {\bibfield  {journal} {\bibinfo  {journal} {Computer Physics
  Communications}\ }\textbf {\bibinfo {volume} {183}},\ \bibinfo {pages} {370}
  (\bibinfo {year} {2012})}\BibitemShut {NoStop}%
\bibitem [{\citenamefont {Scully}\ and\ \citenamefont
  {Zubairy}(1997)}]{Scully_Zubairy_1997}%
  \BibitemOpen
  \bibfield  {author} {\bibinfo {author} {\bibfnamefont {M.~O.}\ \bibnamefont
  {Scully}}\ and\ \bibinfo {author} {\bibfnamefont {M.~S.}\ \bibnamefont
  {Zubairy}},\ }\href@noop {} {\emph {\bibinfo {title} {Quantum Optics}}}\
  (\bibinfo  {publisher} {Cambridge University Press},\ \bibinfo {year}
  {1997})\BibitemShut {NoStop}%
\bibitem [{\citenamefont {Wu}\ \emph {et~al.}(2010)\citenamefont {Wu},
  \citenamefont {Gray},\ and\ \citenamefont {Pelton}}]{wu2010quantum}%
  \BibitemOpen
  \bibfield  {author} {\bibinfo {author} {\bibfnamefont {X.}~\bibnamefont
  {Wu}}, \bibinfo {author} {\bibfnamefont {S.}~\bibnamefont {Gray}},\ and\
  \bibinfo {author} {\bibfnamefont {M.}~\bibnamefont {Pelton}},\ }\bibfield
  {title} {\bibinfo {title} {Quantum-dot-induced transparency in a nanoscale
  plasmonic resonator},\ }\href@noop {} {\bibfield  {journal} {\bibinfo
  {journal} {Optics Express}\ }\textbf {\bibinfo {volume} {18}},\ \bibinfo
  {pages} {23633} (\bibinfo {year} {2010})}\BibitemShut {NoStop}%
\bibitem [{\citenamefont {Yildiz}\ \emph {et~al.}(2020)\citenamefont {Yildiz},
  \citenamefont {Bek},\ and\ \citenamefont {Tasgin}}]{Yildiz2020}%
  \BibitemOpen
  \bibfield  {author} {\bibinfo {author} {\bibfnamefont {B.~C.}\ \bibnamefont
  {Yildiz}}, \bibinfo {author} {\bibfnamefont {A.}~\bibnamefont {Bek}},\ and\
  \bibinfo {author} {\bibfnamefont {M.~E.}\ \bibnamefont {Tasgin}},\ }\bibfield
   {title} {\bibinfo {title} {Plasmon lifetime enhancement in a bright-dark
  mode coupled system},\ }\href@noop {} {\bibfield  {journal} {\bibinfo
  {journal} {Physical Review B}\ }\textbf {\bibinfo {volume} {101}},\ \bibinfo
  {pages} {035416} (\bibinfo {year} {2020})}\BibitemShut {NoStop}%
\bibitem [{\citenamefont {Sahin}(2020)}]{sahin_eot}%
  \BibitemOpen
  \bibfield  {author} {\bibinfo {author} {\bibfnamefont {R.}~\bibnamefont
  {Sahin}},\ }\bibfield  {title} {\bibinfo {title} {Control of eot on
  sub-wavelength au hole arrays via fano resonances},\ }\href
  {https://doi.org/https://doi.org/10.1016/j.optcom.2019.124431} {\bibfield
  {journal} {\bibinfo  {journal} {Optics Communications}\ }\textbf {\bibinfo
  {volume} {454}},\ \bibinfo {pages} {124431} (\bibinfo {year}
  {2020})}\BibitemShut {NoStop}%
\bibitem [{\citenamefont {Günay}\ \emph {et~al.}(2020)\citenamefont {Günay},
  \citenamefont {Artvin}, \citenamefont {Bek},\ and\ \citenamefont
  {Tasgin}}]{Fanoresonance}%
  \BibitemOpen
  \bibfield  {author} {\bibinfo {author} {\bibfnamefont {M.}~\bibnamefont
  {Günay}}, \bibinfo {author} {\bibfnamefont {Z.}~\bibnamefont {Artvin}},
  \bibinfo {author} {\bibfnamefont {A.}~\bibnamefont {Bek}},\ and\ \bibinfo
  {author} {\bibfnamefont {M.~E.}\ \bibnamefont {Tasgin}},\ }\bibfield  {title}
  {\bibinfo {title} {Controlling steady-state second harmonic signal via linear
  and nonlinear fano resonances},\ }\href@noop {} {\bibfield  {journal}
  {\bibinfo  {journal} {Journal of Modern Optics}\ }\textbf {\bibinfo {volume}
  {67}},\ \bibinfo {pages} {26} (\bibinfo {year} {2020})}\BibitemShut {NoStop}%
\bibitem [{\citenamefont {Le~Ru}\ and\ \citenamefont
  {Etchegoin}(2013)}]{ru_QUPERS_figure}%
  \BibitemOpen
  \bibfield  {author} {\bibinfo {author} {\bibfnamefont {E.~C.}\ \bibnamefont
  {Le~Ru}}\ and\ \bibinfo {author} {\bibfnamefont {P.~G.}\ \bibnamefont
  {Etchegoin}},\ }\bibfield  {title} {\bibinfo {title} {Quantifying sers
  enhancements},\ }\href {https://doi.org/10.1557/mrs.2013.158} {\bibfield
  {journal} {\bibinfo  {journal} {MRS Bulletin}\ }\textbf {\bibinfo {volume}
  {38}},\ \bibinfo {pages} {631} (\bibinfo {year} {2013})}\BibitemShut
  {NoStop}%
\end{thebibliography}%

\end{document}


\title{SUPPLEMENTARY MATERIAL: \\
	QUPERS: Quantum Enhanced Raman Spectroscopy}

\author{Özge Demirtas}
\affiliation{{Micro and Nanotechnology Program, Middle East Technical University, 06800 Ankara, Turkey}}
\author{Taner Tarik Aytas}
\affiliation{{Department of Physics, Akdeniz University, 07058 Antalya, Turkey}}
\author{Duygu Gümüs}
\affiliation{{Department of Chemistry, Middle East Technical University, 06800 Ankara, Turkey}}
\author{Deniz Eren Mol}
\affiliation{{Institute of Nuclear Sciences, Hacettepe University, 06100 Ankara, Turkey}}
\affiliation{{Division of Nanotechnology and Nanomedicine, Hacettepe University, 06100 Ankara, Turkey}}
\author{Batuhan Balkan}
\affiliation{{Department of Physics, Middle East Technical University, 06100 Ankara, Turkey}}
\author{Emren Nalbant}
\affiliation{{Department of Chemistry, Middle East Technical University, 06800 Ankara, Turkey}}
\author{Mehmet Emre Tasgin}
\email{metasgin@hacettepe.edu.tr}
\affiliation{{Institute of Nuclear Sciences, Hacettepe University, 06100 Ankara, Turkey}}
\affiliation{{Division of Nanotechnology and Nanomedicine, Hacettepe University, 06100 Ankara, Turkey}}
\author{Ramazan Sahin}
\affiliation{{Department of Physics, Akdeniz University, 07058 Antalya, Turkey}}
\affiliation{{Türkiye National Observatories, TUG, 07058, Antalya, Turkey}}
\author{Alpan Bek}
\email{bek@metu.edu.tr}
\affiliation{{Micro and Nanotechnology Program, Middle East Technical University, 06800 Ankara, Turkey}}
\affiliation{{Department of Physics, Middle East Technical University, 06100 Ankara, Turkey}}

\date{\today}

\maketitle

In this supplementary material, in Sec.~\ref{sec:analytical}, we first demonstrate on a single equation [Eq. (\ref{alpha_R}), below] how a quantum dot~(QD) results path interference effect in the surface enhanced Raman spectrum (SERS) of a metal nanostructure~(MNS)~\cite{postaci_silent_2018}. In Sec.~\ref{sec:Simulations}, we present the details about the simulations with 3D Maxwell equations.  We calculate quantum enhancement factors~(EFs) for different doublets of plasmon modes in Table~\ref{aaa}. 

In Sec.~\ref{2_QD}, we derive an analytical model (solution of Langevin equations) when two quantum dots~(QDs) are attached on a MNS. We demonstrate why Fano enhancement (quantum enhancement, QUPERS) start to degrade. We also demonstrate the same effect via exact solutions of the 3D nonlinear Maxwell equations. In Fig.~\ref{figS3}, we present the quantum enhancement factors~(EFs) for other Raman bands. The EFs display the maximum at QD/AuNS$=1$, similar to Fig.~3a of the main text. in Sec.~\ref{sec:different_Fano}, we describe the difference between Fano enhancement appearing due to lifetime extension and the one taking place at the seady state. The two appear at different frequencies.

\section{Path interference in the nonlinear response} \label{sec:analytical}

In this section, we briefly demonstrate how path interference~(Fano) enhancements take place ---as cancellation in the denominator--- on a simple equation, Eq.~(\ref{alpha_R}). We study the 2 QDs case to Sec.~\ref{2_QD}.

Nonlinear processes~(including SERS) take place over localized intense plasmon modes $\hat{a}$ and $\hat{a}_{\scriptscriptstyle R}$~\cite{grosse2012nonlinear}, of resonances $\Omega$ and $\Omega_{\rm \scriptscriptstyle R}$, respectively. Incident pump ($\omega$) interacts with the Raman molecule~(not to be confused with the QD) strongly. When the Stokes band $\omega_{\scriptscriptstyle R}$ also falls into  one of the localized plasmon modes $\Omega_{\rm \scriptscriptstyle R}$, process possesses very large overlap integrals~\cite{bookchapter,nonLinear_plasmonics_Kauranen2012} yielding very large conversion rates. Yet, $\omega$ and $\omega_{\scriptscriptstyle R}$ may not align with $\Omega$ and $\Omega_{\rm \scriptscriptstyle R}$, respectively. Below, we show that one can still bring the Raman process into resonance utilizing a QO.

Choosing the QD resonance $\Omega_{\rm \scriptscriptstyle QD}$ around $\omega_{\scriptscriptstyle R}$, the Raman amplitude can be determined as
\begin{equation}
	\tilde{\alpha}_{\scriptscriptstyle R}=\frac{-i \chi \varepsilon_{\mathrm{ph}}^{*}}{\beta_{\mathrm{ph}}^{*}\left(\left[i\left(\Omega_{\rm \scriptscriptstyle R}-\omega_{\scriptscriptstyle R}\right)+\gamma_{\scriptscriptstyle R}\right]-\frac{|f|^{2} y}{\left[i\left(\Omega_{\rm  QD}-\omega_{\scriptscriptstyle R}\right)+\gamma_{\rm \scriptscriptstyle QD}\right]}\right)-|\chi|^{2}|\tilde{\alpha}|^{2}} \tilde{\alpha},
	\label{alpha_R}
\end{equation}
using Langevin equations and $c$-numbers~\cite{Premaratne2017}, $\alpha=\langle \hat{a}\rangle$, for the operators $\hat{a}\to\alpha$ and $\hat{a}_{\scriptscriptstyle R}\to \alpha_{\scriptscriptstyle R}$. (For details, see supplementary material in Ref.~\cite{postaci_silent_2018}.) Here, $\chi$ is a constant (overlap integral~\cite{postaci_silent_2018}) determining the strength of the Raman process. $\varepsilon_{\rm ph}$ and $\beta_{\rm ph}$ are factors related with the Raman vibrations. $f$ is the $\hat{a}_{\scriptscriptstyle R} \leftrightarrow$QD coupling strength. $y$ is the population inversion at the steady state. The last term in the denominator $\chi^2 |\alpha|^2$ is extremely small as $\chi$ itself is already too small.

When no QD exists, i.e., $f=0$, the $i(\Omega_{\rm \scriptscriptstyle R}-\omega_{\scriptscriptstyle R}) + \gamma_{\scriptscriptstyle R}$ term determines the Raman conversion factor. This is a usual SERS process, i.e., conversion is maximum (on resonance) when $\omega_{\scriptscriptstyle R}=\Omega_{\rm \scriptscriptstyle R}$. Presence of a QD introduces the extra term $|f|^2y/[i(\Omega_{\rm \scriptscriptstyle QD}-\omega_{\scriptscriptstyle R} ) + \gamma_{\rm \scriptscriptstyle QD}]$. Imaginary part of this term may partially cancel the off-resonant $i(\Omega_{\rm \scriptscriptstyle R}-\omega_{\scriptscriptstyle R})$ term or even bring the process into resonance for certain choices of $i(\Omega_{\rm \scriptscriptstyle QD}-\omega_{\scriptscriptstyle R} ) $.

\begin{figure}
	\includegraphics[width=0.45 \textwidth]{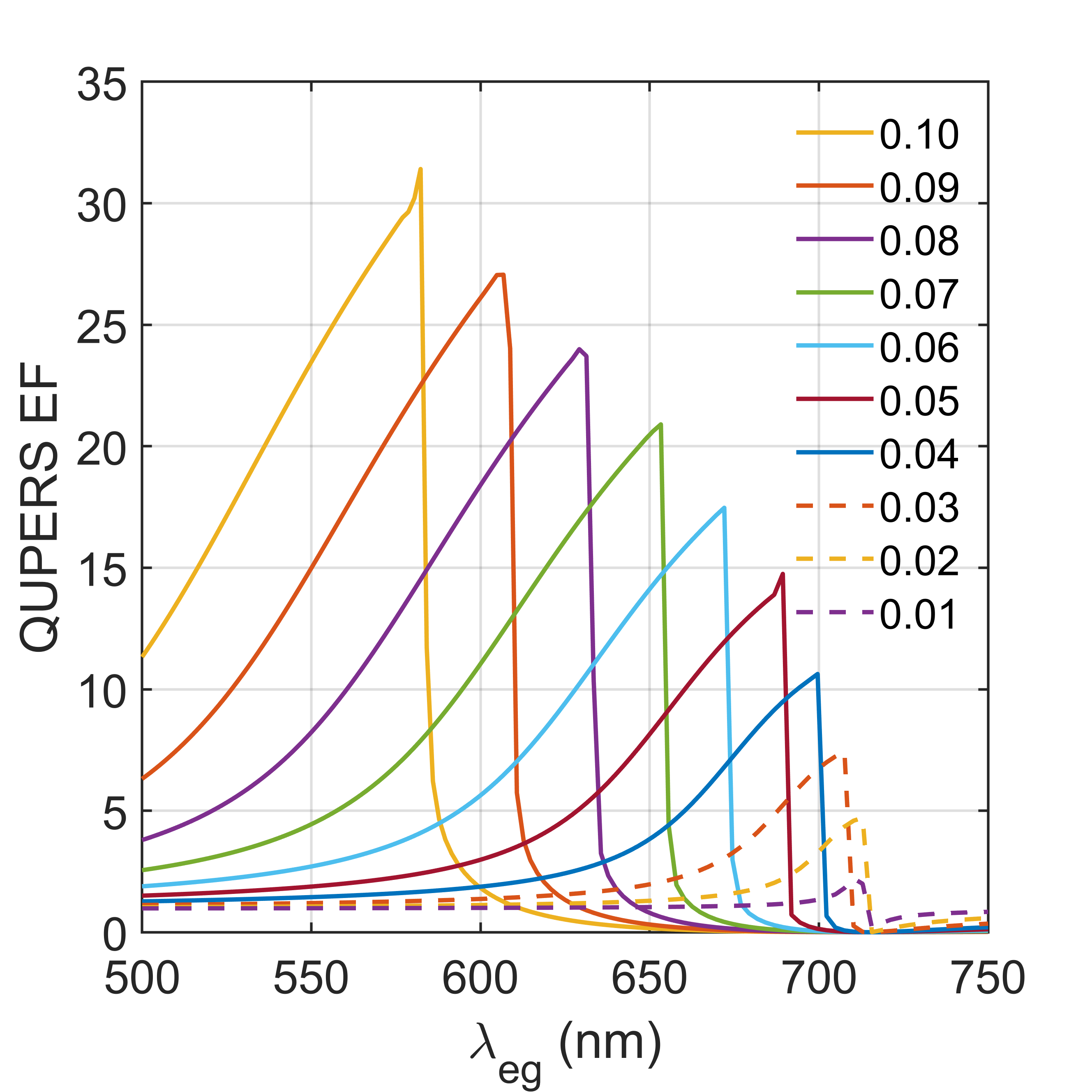}
	\caption{\label{figS1} Quantum enhancement factors (EFs) for different coupling strengths. Different $f$'s (between 0.01 and 0.10) simulate the attachment of a single QD to the AuNS at different positions. When QD is closer to (away from) a hotspot, $f$ is larger (smaller). These EFs are calculated using the experimental parameters. We use $\omega_{\rm exc}=660$ nm, $\omega_{\rm \scriptscriptstyle R}=715$ nm (corresponding to $1165\: {\rm cm}^{-1}$) and $\Omega_{\rm \scriptscriptstyle QD}=$ 562 nm. Here, as an example, we choose the two plasmon modes ($\Omega_1=$655 nm, $\Omega_2$=692 nm). Quantum EFs for other choices of plasmon doublets are given in Table~\ref{aaa}. }
\end{figure}

In Fig.~\ref{figS1}, we observe that Fano enhancement factors of $\sim$10--100 appear for a relatively wide range of choices for different coupling strengths, $f$'s. We use parameters $\omega_{exc}=660$ nm, $\omega_{\rm \scriptscriptstyle R}=715$ nm,  $\Omega_{\rm \scriptscriptstyle QD}=562$ nm, $\gamma_{\rm \scriptscriptstyle QD}=10^{10}$ Hz, $\gamma_{\rm \scriptscriptstyle R}=10^{14}$ Hz, in parallel with our experiments. We obtain the resonances for the plasmon modes as follows. We redraw~(mesh) a nanostar shape from a representative SEM picture. (Note that sizes of synthesized nanostars are stochastic.) We obtain its plasmon spectrum in a freeware program MNPBEM~\cite{hohonester_MNPBEM} and perform a Gaussian fit. Simulations possess several plasmon modes in the regime where Raman process can take place (See Fig.~\ref{figS2}). 

As an example, in Fig.~\ref{figS1}, we use $\Omega=$655 nm and $\Omega_{\rm \scriptscriptstyle R}=$692 nm for two plasmon modes. In fact, nanostar supports many adjacent plasmon modes, see Fig.~\ref{figS2}. So, conversion can take place over many plasmon doublets in our case. Thus, we also perform simulations of the analytical result~\cite{postaci_silent_2018} using other possible doublets. In Table~I, we observe 2 orders of magnitude quantum (Fano) EFs also for other choices of the modes and for different Stokes bands.

Thus, observed EFs can originate only from the Fano effect as ({\it 1}) extra EFs cannot appear due other phenomena as mentioned in the Conclusion section of the main text, ({\it 2}) Fano interference phenomenon predicts extra EFs (multiplying conventional SERS) for a relatively wide ranges of $\Omega_{\rm \scriptscriptstyle QD}$, and ({\it 3}) EFs start to decrease when QD/MNP$>1$ ---a phenomenon can only be observed due to interference.

\section{Simulations} \label{sec:Simulations}

We inquire if the observed quantum EFs also show in the analytical model we have developed for the QUPERS phenomenon in Ref.~\cite{postaci_silent_2018}. Thus, we consider a sample gold nanostar~(AuNS) shape, e.g., given in Fig.~1a of the main text. We perform a triangular meshing on it and simulate its plasmonic response using the freeware toolbox MNPBEM-17~\cite{hohonester_MNPBEM} which is based on boundary element method~(BEM) ---exact solutions of the 3D Maxwell equations. In the simulations, experimental values for the dielectric function of Au~\cite{au_dielectric} are used in the software. The AuNS is illuminated with a ($\lambda_{\rm exc}$) broad-band source polarized along the x-axis. 
 
\begin{figure}[h]
 	\includegraphics[width=0.48 \textwidth]{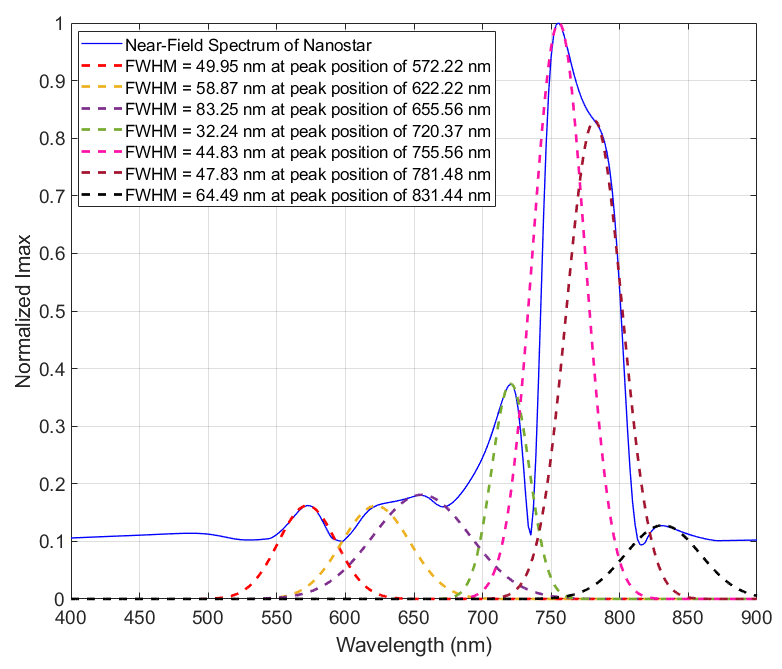}
 	\caption{\label{figS2} Plasmon spectrum of a sample gold nanostar~(AuNS), e.g., depicted in Fig. 1a of the main text. We prepared a 3D mesh for the AuNS and simulated its linear response in freeware toolbox MNPBEM-17~\cite{hohenester2012mnpbem}. We performed fits for the plasmon resonances of the sample AuNS. Quantum enhancement factors are calculated using these plasmon doublets~(see Table~\ref{aaa}) together with the experimental parameters (such as pump and Raman-converted frequencies) in the analytical exact solutions of Langevin equations.}
 \end{figure}

As the Au nanostar is illuminated with a broad-band source, an intense field localization occurs at the hotspots which takes place at or within the spikes.  The plasmon resonances show themselves as sharp peaks or superposition of several such peaks appearing as a relatively broadened response. We sweep the excitation wavelength between 400 nm-900 nm as depicted in Fig.\ref{figS2}. 

We employ the extracted sample plasmon peaks as the input parameters in our exact analytical calculations~\cite{postaci_silent_2018}. We choose several combinations for the $\Omega$ and $\Omega_{\rm \scriptscriptstyle R}$ plasmon modes into which the excitation laser couples and the Raman-converted oscillations take place, respectively. Please note that nonlinear processes take place over the plasmon modes as the overlap integral is extremely large~\cite{grosse2012nonlinear,tasginlinear2018}.  We also determine the damping rates~($\gamma$ and $\gamma_R$) of the modes from the Full Width at Half Maximums (FWHM) we obtain in the fits. Our BEM simulations with the sample AuNS show that individual plasmon modes may cover all the VIS and NIR region of electromagnetic spectrum. In the experiment, UV-VIS absorption spectra carries contribution from different plasmon modes corresponding to different sizes and geometries of Au particles.

 \begin{table}[h]
 	\begin{tabular}{@{}cccccc@{}}
 		\toprule
 		\textbf{$\lambda_{R}$} & \textbf{k $cm^{-1}$} & \textbf{$\Omega$} & \textbf{$\Omega_{R}$} & \textbf{QUPERS EF} & \textbf{f} \\ \midrule
 		675                        & 345                 & 572                & 655                    & 30                 & 0.075      \\
 		675                        & 345                 & 572                & 622                    & 215                & 0.13       \\
 		675                        & 345                 & 572                & 622                    & 140                & 0.125      \\
 		675                        & 345                 & 622                & 655                    & 28                 & 0.075      \\
 		697                        & 805                 & 572                & 655                    & 125                & 0.117      \\
 		697                        & 805                 & 622                & 655                    & 114                & 0.118      \\
 		703                        & 920                 & 622                & 655                    & 97                 & 0.125      \\
 		703                        & 920                 & 572                & 622                    & 18                 & 0.15       \\
 		703                        & 920                 & 655                & 682                    & 28                 & 0.10       \\
 		715                        & 1165                & 622                & 655                    & 245                & 0.15       \\
 		715                        & 1165                & 572                & 655                    & 141                & 0.15       \\
 		715                        & 1165                & 655                & 692                    & 26                 & 0.10       \\
 		726                        & 1371                & 622                & 655                    & 38                 & 0.15       \\
 		726                        & 1371                & 655                & 703                    & 25                 & 0.10       \\
 		737                        & 1581 or 1616        & 622                & 655                    & 11                 & 0.15       \\
 		737                        & 1581 or 1616        & 655                & 714                    & 24                 & 0.10       \\
 		738                        & 1582 or 1616        & 572                & 720                    & 16                 & 0.075      \\ \bottomrule
 	\end{tabular}
 	\caption{Calculated Fano~(quantum) enhancement factors using the exact~(time evolution) solutions of the coupled Langevin equations~\cite{postaci_silent_2018}. These equations, given in Ref.~\cite{postaci_silent_2018}, are the single QD version of the Eqs.~(S3)-(S9) below. We use different combinations of the plasmon mode doublets calculated from the sample spectrum in Fig.~\ref{figS1}. We also use different Raman modes. Laser frequency is $\omega_{\rm exc}$=660 nm as in the experiment.}
 	\label{aaa}
 \end{table}

\section{Two Quantum Dots}\label{2_QD}

In this section, we obtain an analytical model when two QDs attach to the AuNS at different positions. Two different phase factors are assumed to introduce for the two QDs due to the retardation effect. That is, there is a phase delay between the two QDs. We set the delay to $e^{i\theta}$. We derive the hamiltonian and obtain the Langevin equations for such a system. We numerically time-iterate the equation (S3)-(S9) and obtain the steady-state results. We show that  this explains the appearance of the degrading for QD/AuNS$>1$ observed in the experiments. We present the results in Table~\ref{table2}. We also demonstrate the effect using FDTD simulations. This is the affect we observe in the main text and Fig.~\ref{figS3} below.

When two QDs couple to the AuNS at different positions a phase difference $e^{i\theta}$ occurs between the two QDs because of the retardation in the oscillations of the plasmon excitation. Thus, here we extend the model of Ref.~\cite{postaci_silent_2018} merely by coupling an additional QD with a delayed phase factor $e^{i\theta}$ with respect to the coupling of the first QD. The coupling of the second QD~($f_2$) also differs from the first one as it resides at a different position which necessarily affects its interaction with the $\Omega_R$ plasmon mode. Thus, the hamiltonian in Ref.~\cite{postaci_silent_2018} is slightly revised as
\begin{eqnarray} \label{hamiltonian} 
	\hat{\cal H}_{0} & = & \hbar \Omega \hat{a}^{\dagger} \hat{a}+\hbar \Omega_{\mathrm{R}} \hat{a}_{\mathrm{R}}^{\dagger} \hat{a}_{\mathrm{R}}+\hbar \Omega_{\mathrm{ph}} \hat{a}_{\mathrm{ph}}^{\dagger} \hat{a}_{\mathrm{ph}},  \nonumber \\  
	\hat{\cal H}_{\mathrm{QE}}& = & \hbar \omega_{\mathrm{eg_1}}|e\rangle\langle e| \nonumber + \hbar \omega_{\mathrm{eg_2}}|e\rangle\langle e|, \nonumber \\
	\hat{\cal H}_{\mathrm{L}} & = & i \hbar\left(\hat{a}^{\dagger} \varepsilon e^{-i \omega t}-\hat{a} \varepsilon^{*} e^{i \omega t}\right), \\
	\hat{\cal H}_{\mathrm{int}}& = & \hbar\left(f \hat{a}_{\mathrm{R}} \left | e \right\rangle \left\langle g \right  | +f^{*} \hat{a}_{\mathrm{R}}^{\dagger} \left | g \right\rangle \left\langle  e \right |  \right) + \hbar\left(f_{2} e^{-i \theta} \hat{a}_{\mathrm{R}} \left | e \right\rangle \left\langle g \right  | +f^{*}_{2} e^{i \theta} \hat{a}_{\mathrm{R}}^{\dagger} \left | g \right\rangle \left\langle  e \right |  \right), \nonumber \\
	\hat{\cal H}_{\mathrm{R}} & = & \hbar \chi\left(\hat{a}_{\mathrm{R}}^{\dagger} \hat{a}_{\mathrm{ph}}^{\dagger} \hat{a}+\hat{a}^{\dagger} \hat{a}_{\mathrm{ph}} \hat{a}_{\mathrm{R}}\right). \nonumber
\end{eqnarray}

Briefly, the total hamiltonian of the system is the sum of the terms given in Eq.~(S1). $\hat{\cal H}_{0}$ includes energies of the pumped plasmon mode ($\hat{a}$) and plasmon mode into which Raman-converted frequency occurs ($\hat{a}_{\rm \scriptscriptstyle R}$). The Raman vibrations of the molecule is represented by the operator $\hat{a}_{\mathrm{ph}}$ as in Ref.~\cite{postaci_silent_2018}. The QDs (quantum emitters, QE) are modeled as two-level systems~\cite{Scully_Zubairy_1997} where  $|g\rangle$ is the ground and $|e\rangle$ is the excited states of the QE. $\hat{\cal H}_{\rm \scriptscriptstyle {QE}}$ designate the the energy of QE and $\hat{\cal H}_{\mathrm{L}}$ is the coupling of pump to the whole system. $\hat{\cal H}_{\mathrm{ \scriptscriptstyle R}}$ and $\hat{\cal H}_{\mathrm{int}}$ are the Raman conversion process and interaction of the Stokes-shifted plasmon with the QE. 

One obtains the Langevin equations using the Heisenberg equation of motion and replaces the operators with expectation values, i.e., c-numbers~\cite{Premaratne2017}. This way, neglecting quantum entanglement-related effects, one obtains the coupled equations
\begin{eqnarray}
	\dot{\alpha}_{\mathrm{R}} & = & \left(-i \Omega_{\mathrm{R}}-\gamma_{\mathrm{R}}\right) \alpha_{\mathrm{R}}-i \chi \alpha_{\mathrm{ph}}^{*} \alpha-i f^{*} \rho_{\mathrm{ge}}-i f^{*}_{2} e^{i \theta} \rho_{\mathrm{ge_2}}, \\
	\dot{\alpha}& = & (-i \Omega-\gamma) \alpha-i \chi \alpha_{\mathrm{ph}} \alpha_{\mathrm{R}}+\varepsilon e^{-i \omega t},\\
	\dot{\alpha}_{\mathrm{ph}} & = & \left(-i \Omega_{\mathrm{ph}}-\gamma_{\mathrm{ph}}\right) \alpha_{\mathrm{ph}}-i \chi \alpha_{\mathrm{R}}^{*} \alpha+\varepsilon_{\mathrm{ph}} e^{-i \omega_{\mathrm{ph}}{t}}, \\
	\dot{\rho}_{\mathrm{eg}} & = & \left(-i \omega_{\mathrm{eg}}-\gamma_{\mathrm{eg}}\right) \rho_{\mathrm{eg}} {+i f} \alpha_{\mathrm{R}}\left(\rho_{\mathrm{ee}}-\rho_{\mathrm{gg}}\right), \\
	\dot{\rho}_{\mathrm{ee}} & = & -\gamma_{\mathrm{ee}} \rho_{\mathrm{ee}}+i f^{*} \alpha_{\mathrm{R}}^{*} \rho_{\mathrm{eg}}-i f \alpha_{\mathrm{R}} \rho_{\mathrm{eg}}^{*},\\
\dot{\rho}_{\mathrm{eg}_2} & = & \left(-i \omega_{\mathrm{eg}_2}-\gamma_{\mathrm{eg}_2}\right) \rho_{\mathrm{eg}_2} {+i f} \alpha_{\mathrm{R}}\left(\rho_{\mathrm{ee}_2}-\rho_{\mathrm{gg}_2}\right),\\
	\dot{\rho}_{\mathrm{ee}_2} & = & -\gamma_{\mathrm{ee}_2} \rho_{\mathrm{ee}_2}+i f^{*}_2 e^{i\theta} \alpha_{\mathrm{R}}^{*} \rho_{\mathrm{eg}_2}-i f_2 e^{-i\theta} \alpha_{\mathrm{R}} \rho_{\mathrm{eg}_2}^{*}.
\end{eqnarray}
In the supplementary material of Ref.~\cite{postaci_silent_2018}, one can find the details of the derivations for the single QE case.

We numerically time evolve the coupled equations (S3)-(S9) in order to obtain the quantum EFs which are compared to the case without the presence of QEs. In our simulations we scale all frequencies by the pump frequency $\omega_{\rm exc}$=660 nm. We consider a vibrational mode corresponding to 1165 $cm^{-1}$ for the Raman reporter, that is, $\omega_R=715$ nm as in the experiment. The resonance~(absorption) and the decay rate of the QEs are set to ($\Omega_{\rm \scriptscriptstyle QD}$) at 562 nm $\gamma_{\rm \scriptscriptstyle QD}=10^{10}$ Hz~\cite{wu2010quantum}.
For the demonstration of the interference degrading phenomenon, in Table~\ref{table2}, we choose the double plasmon modes from Fig.~\ref{figS1} as $\Omega$=655 nm and $\Omega_{R}$=692 nm.

\begin{table}[h]
	\caption{Degrading of the interference phenomenon for QD/AuNS$>1$. Quantum~(Fano) enhancement factors are calculated from the exact time evolution of Eqs.(S3)-(S9). $\omega_{\rm exc}$=660 nm, $\Omega_{\rm \scriptscriptstyle QD}$=562 nm and $\omega_{\rm \scriptscriptstyle R}$=715 nm (a Raman mode corresponding to 1165 $cm^{-1}$). We take the plasmon modes as $\Omega$=655 nm, $\Omega_{\rm \scriptscriptstyle R}$=692 nm and a fixed coupling strength of $f$=0.15 scaled with $\omega_{\rm exc}$.}
	
	\begin{tabular}{@{}ccc@{}}
		\toprule
		$\theta$                                          & $f_2$ & QUPERS EF  \\ \midrule
		& 0.00 & 144 \\
		& 0.01 & 133 \\
		& 0.05 & 29  \\
		\multirow{-4}{*}{$\pi/2$}                         & 0.10 & 4   \\ \midrule  
		& 0.00 & 144 \\
		& 0.01 & 130 \\
		& 0.05 & 25  \\
		\multirow{-4}{*}{$\pi$/3} & 0.10 & 3   \\ \midrule
		& 0.00 & 144 \\
		& 0.01 & 129 \\
		& 0.05 & 25  \\
		\multirow{-4}{*}{$\pi$/4}                         & 0.10 & 3   \\ \bottomrule
	\end{tabular}\label{table2}
\end{table}

Results in Table \ref{table2} indicate the QUPERS enhancement factors calculated in the presence of two QEs. We present the results for three different phase delays  between the two interactions. We assume that there is no coupling between QEs. $f_2$ is the coupling between AuNS and the second QE. WE clearly see that presence of two QEs at different positions starts to degrade the path interference phenomenon compared to the presence of a single QE. In case both QEs were placed into the same position, i.e., $\theta=0$, however, the quantum EFs would increase.

 \begin{figure}[h]
 	\includegraphics[width=1 \textwidth]{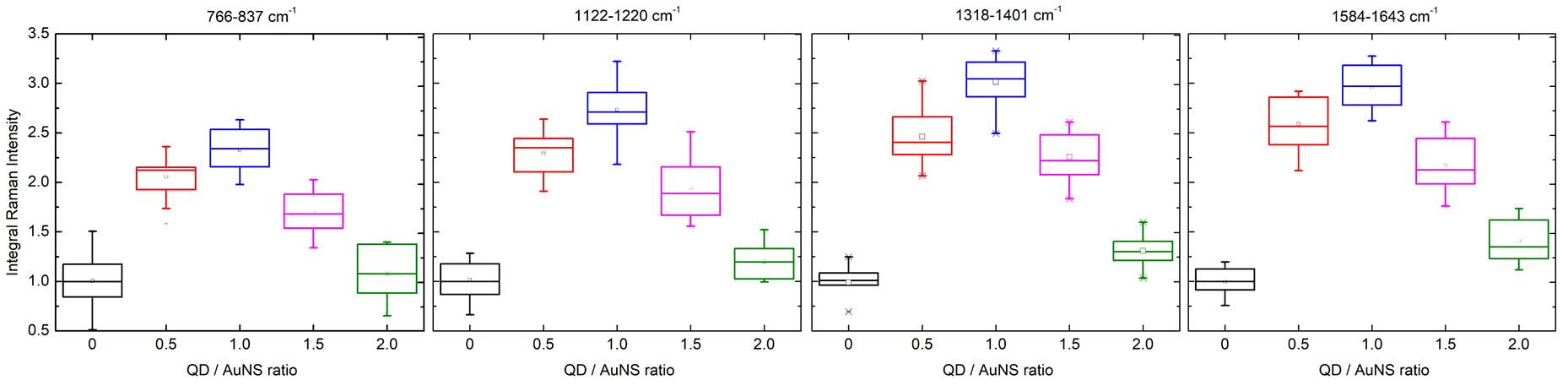}
 	\caption{ Experimental quantum~(Fano) enhancement factors also for different Raman modes, akin to Fig.~3a in the main text. A similar degrading effect appears after QD/AuNS exceeds 1. Occurrence of this effect is confirmed to be owing to the path interference effect. This is shown both using the time evolution of Eqs.~(S3)-(S9), results given in Table~\ref{table2}, and using FDTD simulations depicted in Fig. 3b of the main text. }
 	\label{figS3}
 \end{figure}

In the main text, we present the experimental results with the Raman mode 1165 ${\rm cm}^{-1}$. However, we also measured the quantum EFs for different Raman modes.  Fig.~\ref{figS3} clearly displays that the same phenomenon ---degrading of the quantum EFs for QD/AuNS$>1$--- takes place also for other Raman bands.

We also demonstrated the same degrading effect via exact solutions of 3D nonlinear Maxwell equations. Unfortunately, SERS process is not implemented exactly in neither FDTD~(Lumerical) or FEM~(Comsol) simulation packages. The solvers determine the SERS enhancement factors only approximately. They calculate the field enhancement factors in the input~(excitation, $\omega_{\rm exc}$) and converted~($\omega_{\rm \scriptscriptstyle R}$) frequencies, and simply multiply the two EFs to obtain the SERS enhancement. Still, one can use other ---such as second harmonic or third harmonic generation~(THG)--- processes which are held exactly in the time dependent Maxwell equations.  Thus, we showed the interference-degrading phenomenon in the THG process using Lumerical FDTD tool. In Fig.~3b one clearly observes that when we put more than a single QD~(modeled by Lorentzian dielectric function~\cite{wu2010quantum}), degrading of the enhancement due to the path interference starts. This is the effect we also observe from the analytical solutions of Eq.~(S3)-(S9) whose results are depicted in Table~\ref{table2}.

\section{Different Types of Fano Effects} \label{sec:different_Fano}
 
 Coupling of a long excitation lifetime mode---usually dark modes are employed in the literature~\cite{Yildiz2020}--- to bright modes introduces three kinds of effects. (i) When one uses an ultrashort laser as the source, due to their high intensities, occupancy of the bright modes becomes several orders larger owing to the extension of their lifetimes~\cite{Yildiz2020}. (ii) When one uses a CW source,  one finds the long-time (steady state) behavior. In this case, the reverse phenomenon takes place: an excitation dip appears at $\omega=\Omega_{\rm \scriptscriptstyle QD}$~\cite{bookchapter,sahin_eot}. Such a dip takes place both in the linear and nonlinear response \cite{,Fanoresonance}. (iii) In the steady-state, for careful choices of the $\Omega_{\rm \scriptscriptstyle QD}$, nonlinear process can be brought into resonance (enhanced). This is the case we study here where cancellation in the denominator of the Raman amplitude $\alpha_{\scriptscriptstyle R}$ takes place in Eq.~(\ref{alpha_R}) above. In our experiment and analytical model, coupling of QDs causes the cancellation in the denominator, so the enhancement effect. 

 \begin{figure}[h]
 	\includegraphics[width=0.65 \textwidth]{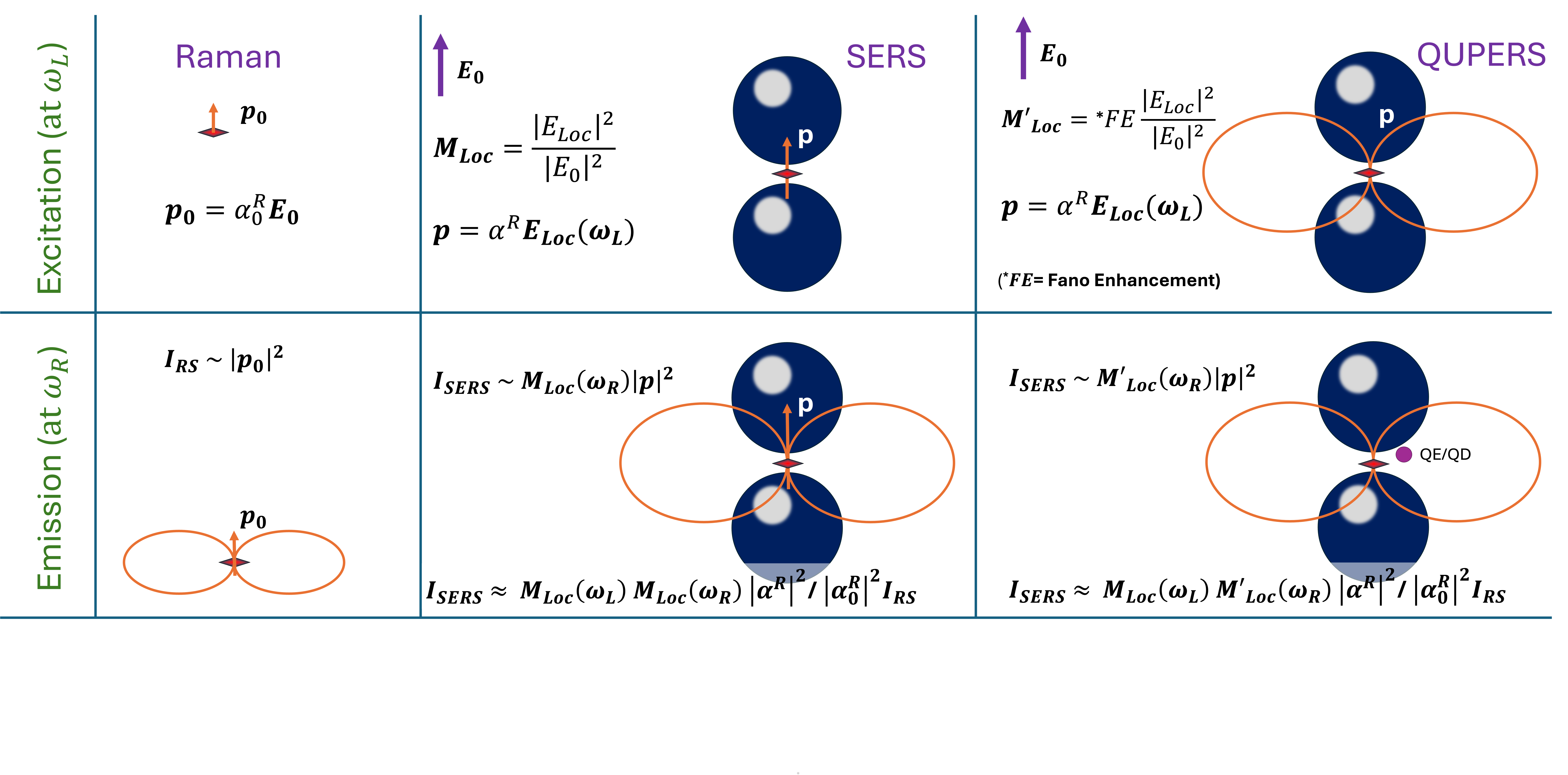}
 	\caption{\label{figS4} Schematic illustration of the origin of the surface-enhanced Raman spectroscopy (SERS) enhancement. The first column indicates the normal Raman scattering process, the middle column indicates the enhancement in excitation experienced by a dipole in a strong local field $E_{Loc}$ (quantified by $M_{Loc}$), together with a modified polarizability $\alpha_{R}$ that includes any possible chemical effect. The third column indicates the QUPERS enhancement phenomena.}
 \end{figure}

 The Fig.~\ref{figS4} depicts the normal Raman, SERS and QUPERS schemes comperatively regarding the excitation and emission intensities and provides the formalisms in the context of chemical and physical enhancement channels ~\cite{ru_QUPERS_figure}.
 
\bibliography{bibliography2}